\documentclass[aps,prd,superscriptaddress,twocolumn]{revtex4-1}
\usepackage{amssymb}
\usepackage{amsmath}
\usepackage{wallpaper}
\usepackage{bm}
\usepackage{multirow}
\usepackage[colorlinks,linkcolor=blue, urlcolor=blue, anchorcolor=blue, citecolor=blue]{hyperref}

\begin{document}

\title{Properties and microscopic structures of dense stellar matter in RMF models}

\author{Jia-Xing~Niu}
\affiliation{Center for Gravitation and Cosmology, College of Physical Science and Technology, Yangzhou University, Yangzhou 225009, China}

\author{Hao Sun}
\affiliation{Center for Gravitation and Cosmology, College of Physical Science and Technology, Yangzhou University, Yangzhou 225009, China}

\author{Cheng-Jun Xia}
\email{cjxia@yzu.edu.cn}
\affiliation{Center for Gravitation and Cosmology, College of Physical Science and Technology, Yangzhou University, Yangzhou 225009, China}

\author{Toshiki Maruyama}
\affiliation{Advanced Science Research Center, Japan Atomic Energy Agency, Shirakata 2-4, Tokai, Ibaraki 319-1195, Japan}

\date{\today}

\begin{abstract}
Data tables on the equation of state (EOS) and microscopic structures for cold dense stellar matter with proton fractions $Y_p =0.01$-$0.65$ and baryon number densities $n_\text{b}=10^{-8}$-$2 \ \mathrm{fm}^{-3}$ are obtained adopting 13 different relativistic density functionals, i.e., NL3, PK1, PK1r, GM1, MTVTC, DD-LZ1, PKDD, DD-ME2, TW99, DD-MEX, DD-MEX1, DD-MEX2, and DD-MEY. The EOSs of dense stellar matter inside neutron stars with baryon number densities $n_\text{b}=7.6\times 10^{-11}$-$2 \ \mathrm{fm}^{-3}$ are obtained as well fulfilling $\beta$-stability condition. In general, the dense stellar matter exhibits droplet phase at $n_\mathrm{b}\lesssim 0.015\ \mathrm{fm}^{-3}$, while more exotic structures such as rods, slabs, tubes, and bubbles appear sequentially as density increases. The critical proton fractions $Y_p^\mathrm{drip}$ ($\approx 0.26$-0.31) for neutron drip are obtained, where neutron gas emerges outside of nuclei at $Y_p< Y_p^\mathrm{drip}$. For dense stellar matter at small densities ($n_\text{b}\lesssim 10^{-5} \ \mathrm{fm}^{-3}$) or large proton fractions ($n_\text{b}\lesssim0.1 \ \mathrm{fm}^{-3}$ and $Y_p\gtrsim Y_p^\mathrm{drip}$), the EOSs and microscopic structures are generally insensitive to the adopted density functionals. With the onset of neutron drip at $Y_p\lesssim Y_p^\mathrm{drip}$, the uncertainties emerge and peak at $n_\text{b} \approx 0.02 \ \mathrm{fm}^{-3}$ within the range $10^{-5} \lesssim n_\text{b}\lesssim0.1 \ \mathrm{fm}^{-3}$. At $n_\text{b}\gtrsim0.1 \ \mathrm{fm}^{-3}$, the dense stellar matter becomes uniform and muons eventually appear, where the uncertainties in the EOSs grow significantly.
\end{abstract}

\maketitle

\section{\label{sec:intro}Introduction}

The equation of state (EOS) is an essential ingredient in understanding the evolution of dense stellar matter associated with a variety of astronomical phenomena, e.g., the core-collapse supernova, evolutions of neutron stars, and binary neutron star mergers~\cite{Pons1999_ApJ513-780, Horowitz2004_PRC69-045804, Lattimer2012_ARNPS62-485, Janka2012_ARNPS62-407, Bauswein2012_PRD86-063001, Rueda2014_PRC89-035804, Qi2016_RAA16-008, Watanabe2017_PRL119-062701, Sotani2019_MNRAS489-3022, Koeppel2019_ApJ872-L16, Baiotti2019_PPNP109-103714, Schuetrumpf2020_PRC101-055804, Bauswein2020_PRL125-141103, Gittins2020_PRD101-103025, Preau2021_MNRAS505-939}. Due to the difficulties in simulating dense matter with lattice QCD, it is challenging to obtain unambiguously the EOS of dense stellar matter throughout the large density and temperature ranges covered in various astronomical phenomena.

The properties of nuclear matter around the saturation density ($n_0\approx 0.16 \ \mathrm{fm}^{-3}$), nevertheless, are well constrained by various experimental and theoretical efforts, where the binding energy $B\approx -16$ MeV, the incompressibility $K = 240 \pm 20$ MeV~\cite{Shlomo2006_EPJA30-23}, the symmetry energy $ S = 31.7 \pm3.2$ MeV and its slope $L = 58.7\pm28.1$ MeV~\cite{Li2013_PLB727-276, Oertel2017_RMP89-015007}. More recent measurements of the neutron-skin thicknesses from PREX-II ($^{208}$Pb)~\cite{PREX2021_PRL126-172502} and CREX ($^{48}$Ca)~\cite{CREX2022_PRL129-042501} are expected to update the slope of the symmetry energy, which nonetheless require more investigations as the slopes constrained by the two experiments turn out to be in strong tension with each other~\cite{Zhang2020_PRC101-034303, Essick2021_PRL127-192701, Reinhard2022_PRL129-232501}. Further constraints on nuclear matter at larger densities ($\lesssim 4.5n_0$) can be attained via relativistic heavy-ion collisions~\cite{Danielewicz2002_Science298-1592, Liu2021_PRC103-014616}, e.g., the skewness coefficient $J=-390^{+60}_{-70}$ MeV if the incompressibility has a uniform prior probability distribution function in the range of $K = 240 \pm 20$ MeV~\cite{Xie2021_JPG48-025110}.

Meanwhile, as we are in the era of multimessenger astronomy, the structures and evolutions of compact stars provide further constraints on the properties of dense stellar matter.
For example, by carefully analyzing the arriving times of radio pulses from pulsars in a binary system, their masses can be precisely measured, e.g., the two-solar-mass pulsars PSR J1614-2230 ($1.928 \pm 0.017\ M_\odot$)~\cite{Demorest2010_Nature467-1081, Fonseca2016_ApJ832-167} and PSR J0348+0432 ($2.01 \pm 0.04\ M_\odot$)~\cite{Antoniadis2013_Science340-1233232}. Carrying out pulse-profile modelings of X-ray pulses originated from the hot spots of pulsars, their masses and radii can be measured, e.g., PSR J0030+0451~\cite{Riley2019_ApJ887-L21, Miller2019_ApJ887-L24}, PSR J0740+6620~\cite{Riley2021_ApJ918-L27, Miller2021_ApJ918-L28}, PSR J0437-4715~\cite{Choudhury2024_ApJ971-L20}, and PSR J1231-1411~\cite{Salmi2024_ApJ976-58}. The gravitational and electromagnetic signals originated from the binary compact star merger events {GRB} 170817A-{GW}170817-{AT} 2017gfo, GW190814, and GW200210~\cite{LVC2018_PRL121-161101, LVC2020_ApJ896-L44, Zhu2022_ApJ928-167} could also provide various information on constrianing the masses, tidal deformabilities, and radii of neutron stars.

Additionally, the upcoming observations with the LIGO/Virgo/KAGRA detector network, the Cosmic Explorer~\cite{Reitze2019_BAAS51-035}, the Einstein Telescope~\cite{Punturo2010_CQG27-194002}, and the Neutron Star Extreme Matter Observatory (NEMO)~\cite{Ackley2020_PASA37-e047} are expected to have increased sensitivity. Combined with various electromagnetic and neutrino signals, the dynamical evolution of core-collapse supernovae and the protoneutron stars could help to constrain the EOS of dense stellar matter at large densities and temperatures~\cite{Mueller2013_ApJ766-43, Kawahara2018_ApJ867-126, Sotani2022_EPJC82-477, Sotani2022_PRD105-023007, Zha2024_PRD109-083023, Mezzacappa2024}. The improved high frequency sensitivity of the gravitational wave detectors would enable us to measure the gravitational waves emitted during and after the merger of neutron stars, which provide probes to identify a possible deconfinement phase transition at highest densities~\cite{Bauswein2019_PRL122-061102, Huang2022_PRL129-181101, Fujimoto2023_PRL130-91404} as well as the out-of-equilibrium effects in compact stars~\cite{Most2021_MNRAS509-1096, Most2024_AJL967-14, Zappa2023_MNRAS520-1481}.

In such cases, it is necessary to provide the EOSs covering the full range of densities, proton fractions, and temperatures involved in various astronomical environments~\cite{Baym1971_ApJ170-299, Shen2011_PRC83-035802, Shen2011_ApJ197-20, Togashi2017_NPA961-78, Oertel2017_RMP89-015007}. Considering the uncertainty of nuclear energy density functionals, various EOSs predicted by those functionals enclosing the uncertainty band are needed, which could be further constrained based on the numerical simulations of supernova explosions, binary neutron star merges, asteroseismology of neutron stars, as well as the evolution of various neutron star properties, e.g., temperature, magnetic field, and spin frequencies. In particular, at subsaturation densities, a first-order liquid-gas phase transition takes place, forming various nonuniform structures of nuclear matter, i.e., nuclear pasta~\cite{Baym1971_ApJ170-299, Negele1973_NPA207-298, Ravenhall1983_PRL50-2066, Hashimoto1984_PTP71-320, Williams1985_NPA435-844}, which play important roles in the transport and elastic properties of neutron stars~\cite{Chamel2008_LRR11-10, Caplan2017_RMP89-041002, Zhu2023_PRD107-83023, Sotani2024_Universe10-231}.

Adopting spherical and cylindrical approximations for the Wigner-Seitz (WS) cells~\cite{Pethick1998_PLB427-7, Oyamatsu1993_NPA561-431, Maruyama2005_PRC72-015802, Togashi2017_NPA961-78, Shen2011_ApJ197-20}, we can then fix the geometrical structures of nuclear pasta. At a given average density $n_\mathrm{b}$, the optimal sizes of the WS cells are fixed by minimizing the energy of dense stellar matter. As the average density increases, five distinct structural types emerge in sequence, i.e., droplets, rods, slabs, tubes, and bubbles, which ultimately becomes uniform at large enough densities and temperatures. In the present study, we restrict ourselves to cold dense stellar matter ($T=0$), therefore the EOS depend on two state variables, i.e., the baryon number density ($n_\text{b}$) and proton fraction ($Y_p$). To avoid introducing additional uncertainties~\cite{Fortin2016_PRC94-035804, DinhThi2021_AA654-A114}, in this work we investigate the EOSs and microscopic structures of dense stellar matter in a unified manner~\cite{Xia2022_CTP74-095303}, where the Thomas-Fermi approximation (TFA) is employed. In particular, similar to Refs.~\cite{Maruyama2005_PRC72-015802, Avancini2008_PRC78-015802, Avancini2009_PRC79-035804, Gupta2013_PRC87-028801}, the properties of nuclear matter are fixed with relativistic mean field (RMF) models~\cite{Meng2016_RDFNS}, which well describes the properties of finite nuclei~\cite{Reinhard1989_RPP52-439, Ring1996_PPNP37_193-263, Meng2006_PPNP57-470, Paar2007_RPP70-691, Meng2015_JPG42-093101, Meng2016_RDFNS, Chen2021_SCPMA64-282011, Typel1999_NPA656-331, Vretenar1998_PRC57-R1060, Lu2011_PRC84-014328} and nuclear matter~\cite{Glendenning2000, Ban2004_PRC69-045805, Weber2007_PPNP59-94, Long2012_PRC85-025806, Sun2012_PRC86-014305, Wang2014_PRC90-055801, Fedoseew2015_PRC91-034307, Gao2017_ApJ849-19}. Our calculation employs 13 different relativistic density functionals (RDFs) including both nonlinear self-couplings (NL3~\cite{Lalazissis1997_PRC55-540}, PK1~\cite{Long2004_PRC69-034319},  PK1r~\cite{Long2004_PRC69-034319}, GM1~\cite{Glendenning1991_PRL67-2414}, MTVTC~\cite{Maruyama2005_PRC72-015802}) and density-dependent couplings (DD-LZ1~\cite{Wei2020_CPC44-074107}, PKDD~\cite{Long2004_PRC69-034319}, DD-ME2~\cite{Lalazissis2005_PRC71-024312}, TW99~\cite{Typel1999_NPA656-331}, DD-MEX~\cite{Taninah2020_PLB800-135065}, DD-MEX1, DD-MEX2, and DD-MEY~\cite{Taninah2023_PRC107-041301}), which are fixed by reproducing finite nuclei properties and fulfill the two-solar-mass constraint for neutron stars.

In this work, based on TFA and adopting spherical and cylindrical approximations for WS cells, we fix the mean fields and density profiles iteratively until convergency is reached. The reflective boundary condition for the mean fields are fulfilled by expanding them with cosine functions based on fast cosine transformation, where the charge screening effects due to the relocation of charged particles are considered self-consistently. The nuclear pastas are then fixed self-consistently by searching for the optimum configuration that minimizes the energy of the system at fixed average density $n_\text{b}$ and proton fraction $Y_p$. The EOS tables for cold dense stellar matter as functions of $n_\text{b}$ and $Y_p$ obtained with the 13 RDFs are then provided.

The paper is organized as follows. In Sec.~\ref{sec:Theor} we present the theoretical framework of RMF models and the numerical details on fixing the microscopic structures of dense stellar matter. The obtained results and the corresponding discussions are then illustrated in Sec.~\ref{sec:res}. Our conclusion is presented in Sec.~\ref{sec:con}.

\section{\label{sec:Theor}Theoretical framework}
\subsection{\label{sec:rmf}RMF models}

The Lagrangian density of RMF model for dense stellar matter is expressed as
 \begin{align}
 \mathcal{L}=&\sum_{i=n,p}\bar\psi_i[i\gamma^\mu\partial_\mu-\gamma^0(\textsl{g}_{\omega}\omega+\textsl{g}_{\rho}\rho\tau_i+Aq_i)-m_i^{*}]\psi_i\nonumber\\
 &+\sum_{l=e,\mu}\bar\psi_l[i\gamma^\mu\partial_\mu-m_l+e\gamma^0A]\psi_l-\frac{1}{4}A_{\mu\nu}A^{\mu\nu}\nonumber\\
 &+\frac{1}{2}\partial_\mu\sigma\partial^\mu\sigma-\frac{1}{2}m_\sigma^2\sigma^2-\frac{1}{4}\omega_{\mu\nu}\omega^{\mu\nu}+\frac{1}{2}m_\omega^2\omega^2\nonumber\\
 &-\frac{1}{4}\rho_{\mu\nu}\rho^{\mu\nu}+\frac{1}{2}m_\rho^2\rho^2+U(\sigma,\omega, \rho), \label{equ-1}
 \end{align}
where ${\tau}_n=-{\tau}_p=1$ represents the third component of isospin for nucleons, $q_p=-q_e=-q_\mu=e$ and $q_n=0$ the charge, and $m^*_{n,p}\equiv m_{n,p}+ \textsl{g}_\sigma\sigma$ the effective nucleon mass. Because of time reversal symmetry, the boson fields $\sigma$, $\omega$, $\rho$ and $A$ take mean values and are left with temporal parts. So that the field tensors $\omega_{\mu\nu}$, $\rho_{\mu\nu}$ and $A_{\mu\nu}$ vanish except for
\begin{equation}
\omega_{i0}=-\omega_{0i}=\partial_i\omega,\rho_{i0}=-\rho_{0i}=\partial_i\rho, A_{i0}=-A_{0i}=\partial_iA. \nonumber
\end{equation}
The nonlinear self couplings of the mesons are determined by
\begin{equation}\label{equ-2}
    U(\sigma,\omega, \rho)=-\frac{1}{3}\textsl{g}_2\sigma^3-\frac{1}{4} \textsl{g}_3\sigma^4+\frac{1}{4}c_3\omega^4+\frac{1}{4}d_3\rho^4,
\end{equation}
while the density-dependent coupling constants according to the Typel-Wolter ansatz~\cite{Typel1999_NPA656-331} are expressed as
\begin{align}
    \label{equ-3}
    \textsl{g}_\xi(n_\text{b})& =\textsl{g}_\xi a_\xi\frac{1+b_\xi(n_\text{b}/n_0 + d_\xi)^2}{1+c_\xi(n_\text{b}/n_0 + d_\xi)^2},\\
   \label{equ-4}
    \textsl{g}_\rho(n_\text{b})&=\textsl{g}_\rho \text{exp}[-a_\rho(n_\text{b}/n_0-1)],
\end{align}
where $\xi=\sigma$ and $\omega$. Note that the coefficients in Eq.~(\ref{equ-2}) vanish with $\textsl{g}_2=\textsl{g}_3=c_3=d_3=0$ if we adopt the density-dependent RDFs mentioned above. For completeness, in Table~\ref{table:param} we present the adopted parameter sets, where $a_{\sigma, \omega}=1$ and $b_{\sigma, \omega}=c_{\sigma, \omega}=a_\rho=0$ if nonlinear self-couplings are adopted.

\begin{table*}
\caption{\label{table:param} The adopted parameters for the RMF models with nonlinear self-couplings (NL3~\cite{Lalazissis1997_PRC55-540}, PK1~\cite{Long2004_PRC69-034319},  PK1r~\cite{Long2004_PRC69-034319}, GM1~\cite{Glendenning1991_PRL67-2414}, MTVTC~\cite{Maruyama2005_PRC72-015802}) and density-dependent couplings (DD-LZ1~\cite{Wei2020_CPC44-074107}, PKDD~\cite{Long2004_PRC69-034319}, DD-ME2~\cite{Lalazissis2005_PRC71-024312}, TW99~\cite{Typel1999_NPA656-331}, DD-MEX~\cite{Taninah2020_PLB800-135065}, DD-MEX1, DD-MEX2, and DD-MEY~\cite{Taninah2023_PRC107-041301}).}
\begin{tabular}{c|cccccccc|cccc} \hline \hline
       & $m_n$   & $m_p$   & $m_\sigma$& $m_\omega$ & $m_\rho$ & $g_\sigma$  & $g_\omega$ & $g_\rho$  &  $g_2$      &   $g_3$   & $c_3$ & $d_3$ \\
       &   MeV   &   MeV   &   MeV     &      MeV   &   MeV    &             &            &           & fm${}^{-1}$ &           &        \\ \hline
NL3    & 939     &   939   &  508.1941 &    782.501 &    763   &  10.2169    &  12.8675   & 4.4744    &  $-$10.4307 & $-$28.8851& 0 & 0     \\
PK1    & 938     &   938   &  511.198  &    783     &    770   &  10.0289    &  12.6139   & 4.6322    &  $-$7.2325  &    0.6183 & 71.3075& 0\\
PK1r    & 939.5731     &   938.2796   &  514.0873  &    784.222     &    763   &  10.3219    &  13.0134   &  4.55    &  $-$8.1562  &    10.1984 & 54.4459& 350\\
GM1    & 938     &   938   &  510      &    783     &    770   &  8.87443    &  10.60957  & 4.09772   &  $-$9.7908  & $-$6.63661& 0 & 0     \\
MTVTC  & 938     &   938   &  400      &    783     &    769   &  6.3935     &  8.7207    & 4.2696    &  $-$10.7572 & $-$4.04529& 0 & 0     \\ \hline
DD-LZ1 & 938.9   &  938.9  &538.619216 &    783     &    769   &  12.001429  &  14.292525 & 3.486227  &       0     &    0      & 0 & 0 \\
DD-MEX & 938.5   &  938.5  &547.332728 &    783     &    763   &  10.706722  &  13.338846 & 3.619020  &       0     &    0      & 0 & 0 \\
DD-MEX1 & 939   &  939  & 553.714785 &    783    &    763   &  10.668226  &  13.107751 & 3.641508  &       0     &    0      & 0 & 0 \\
DD-MEX2 & 939   &  939 &551.087886 &    783     &    763   &  10.476976  &  12.903532 & 4.100719  &       0     &    0      & 0 & 0 \\
DD-MEY & 939  &  939  &551.321796 &    783     &    763   &  10.411867 &  12.803298 & 3.69217  &       0     &    0      & 0 & 0 \\
PKDD   &939.5731 &938.2796 &555.5112   &    783     &    763   &  10.7385    &  13.1476   & 4.2998    &       0     &    0      & 0 & 0 \\
DD-ME2 & 938.5   &  938.5  &550.1238   &    783     &    763   &  10.5396    &  13.0189   & 3.6836    &       0     &    0      & 0 & 0 \\
TW99   & 939     &  939    &  550      &    783     &    763   &  10.7285    &  13.2902   & 3.6610    &       0     &    0      & 0  & 0\\
\hline
\end{tabular}

\begin{tabular}{c|cccc|cccc|c} \hline \hline
       & $a_\sigma$&$b_\sigma$&$c_\sigma$&$d_\sigma$ &$a_\omega$& $b_\omega$ & $c_\omega$ & $d_\omega$ & $a_\rho$   \\ \hline
DD-LZ1 &1.062748 &1.763627 &2.308928 & 0.379957 &1.059181 &0.418273 &0.538663 &0.786649 &0.776095  \\
DD-MEX &1.397043 &1.334964 &2.067122 & 0.401565 &1.393601 &1.019082 &1.605966 &0.455586 &0.620220  \\
DD-MEX1 &1.392047 &2.107233 &3.156692& 0.324955 &1.382434&1.880412 &2.811153&0.344348 & 0.561222  \\
DD-MEX2 &1.40869 &1.50646 &2.337477 & 0.377629 &1.404071 &1.349038 &2.100795 &0.398334 & 0.19354 \\
DD-MEY &1.4372 &2.059712 &3.210289 & 0.322231 &1.431375 &1.943724 &3.025356 &0.331934 &0.532267  \\
PKDD   &1.327423 &0.435126 &0.691666 & 0.694210 &1.342170 &0.371167 &0.611397 &0.738376 &0.183305  \\
DD-ME2 &1.3881   &1.0943   &1.7057   & 0.4421   &1.3892   &0.9240   & 1.4620  &0.4775   &0.5647    \\
TW99   &1.365469 &0.226061 &0.409704 & 0.901995 &1.402488 &0.172577 &0.344293 &0.983955 &0.515000  \\
\hline
\end{tabular}
\end{table*}

Based on the variational principle, the equations of motion for boson fields are fixed by
\begin{align}
\label{equ-5}    (-\nabla^2 + m^2_\sigma)\sigma &=-\textsl{g}_\sigma n_s-\textsl{g}_2\sigma^2-\textsl{g}_3\sigma^3,\\
\label{equ-6}    (-\nabla^2 + m^2_\omega)\omega &=\textsl{g}_\omega n_\text{b}-c_3\omega^3,\\
\label{equ-7}    (-\nabla^2 + m^2_\rho)\rho &=\sum _{i=n,p}\textsl{g}_\rho\tau_i n_i -d_3\rho^3,\\
\label{equ-8}    -\nabla^2A &=e(n_p -n_e-n_\mu).
\end{align}
In this paper, we consider only zero-temperature cases and adopt no sea approximation, the scalar and vector densities are then determined by
\begin{align}
\label{equ-9}    n_s &= \sum_{i=n,p}\langle\bar\psi_i\psi_i\rangle= \sum_{i=n,p}\frac{M^{*3}}{2\pi^2}\textsl{g}\left(\frac{\nu_i}{M^*}\right),\\
\label{equ-10}    n_i & =\langle\bar\psi_i\gamma^0\psi_i\rangle=\frac{\nu^3_i}{3\pi^2},
\end{align}
with $\nu_i$ being Fermi momentum and $\textsl{g}(x)=x\sqrt{x^2+1}-\text{arcsh}(x)$. The total energy of system is given by
\begin{equation}
\label{equ-11}    E=\int \langle\mathcal{T}_{00}\rangle d^3r,
\end{equation}
where the energy momentum tensor is obtained with
\begin{align}
    \langle\mathcal{T}_{00}\rangle =&\sum_{i}\frac{m^{*4}_i}{8\pi^2}\left[x_i(2x^2_i+1)\sqrt{x^2_i+1}-\text{arcsh}(x_i)\right] \nonumber\\
    &+ \frac{1}{2}(\nabla\sigma)^2+\frac{1}{2}m_\sigma^2\sigma^2+\frac{1}{2}(\nabla\omega)^2+\frac{1}{2}m_\omega^2\omega^2 \nonumber\\
    &+\frac{1}{2}(\nabla\rho)^2+\frac{1}{2}m_\rho^2\rho^2 +\frac{1}{2}(\nabla A)^2 \nonumber\\
    &+c_3\omega^4+d_3\rho^4-U(\sigma,\omega, \rho). \label{equ-12}
\end{align}
Here the first term represents the local kinetic energy density, the quantity $x_i\equiv \nu_i/m^*_i$, the lepton masses $m^*_e=m_e=0.511$ MeV and $m^*_\mu=m_\mu=105.66$ MeV.
For any given density profiles $n_i(\Vec{r})$, the total particle numbers $N_i=\int n_id^3r$ and energy $E$ of the system can be obtained. Then we fix the ground state by minimizing $E$ with respect to the density profiles $n_i(\Vec{r})$ at fixed $N_i$, which follows the constancy of chemical potentials, i.e.
\begin{equation}\label{equ-13}
\mu_i(\Vec{r})=\sqrt{\nu^2_i+m^{*2}_i}+\Sigma^R+\textsl{g}_\omega\omega+\textsl{g}_\rho\tau_i\rho+q_iA=\mathrm{constant},
\end{equation}
with the additional ``rearrangement" term
\begin{equation}\label{equ-14}
    \Sigma^R = \frac{d\textsl{g}_\sigma}{dn_\text{b}}\sigma n_s+\frac{d\textsl{g}_\omega}{dn_\text{b}}\omega n_\text{b} +\frac{d\textsl{g}_\rho}{dn_\text{b}}\rho \sum_i\tau_i n_i
\end{equation}
arises from the density-dependent coupling constants~\cite{Lenske1995_PLB345-355}.

\subsection{\label{sec:num}Numerical Details}
By solving equations~(\ref{equ-5})-(\ref{equ-8}) inside a WS cell with the density distributions fixed by fulfilling Eq.~(\ref{equ-13}), microscopic structures of dense stellar matter can be obtained in an iterative manner, where the imaginary time step method is adopted to fix the density profiles during each iteration~\cite{Levit1984_PLB139-147}. In practice, we adopt the spherical and cylindrical approximations, so that the mean field equations~(\ref{equ-5})-(\ref{equ-8}) are reduced to one-dimensional, i.e.,
\begin{align}
\label{equ-15}    1\text{D}: \nabla^2\phi(\Vec{r})&=\frac{\mbox{d}^2\phi(r)}{\mbox{d}r^2},\\
\label{equ-16}    2\text{D}: \nabla^2\phi(\Vec{r})&=\frac{\mbox{d}^2\phi(r)}{\mbox{d}r^2}+\frac{1}{r}\frac{\mbox{d}\phi(r)}{\mbox{d}r},\\
\label{equ-17}    3\text{D}: \nabla^2\phi(\Vec{r})&=\frac{\mbox{d}^2\phi(r)}{\mbox{d}r^2}+\frac{2}{r}\frac{\mbox{d}\phi(r)}{\mbox{d}r},
\end{align}
with $\phi=\sigma$, $\omega$, $\rho$, and $A$. Those differential equations are solved with fast cosine transformation fulfilling reflective boundary conditions ${\mbox{d}\phi(r)}/{\mbox{d}r}=0$ at $r=0$ and $r=R_\mathrm{W}$ as illustrated in Ref.~\cite{Xia2021_PRC103-055812}. To check if convergency is reached, the deviation of local chemical potentials fixed by
\begin{equation}
\label{equ-18}    \sum_{i=p,n,e,\mu}\langle\Delta\mu^2_i\rangle=\sum_{i=p,n,e,\mu}\frac{\int [\mu_i(\Vec{r})-\bar\mu_i]^2n_i(\Vec{r})d^3r}{\int n_i(\Vec{r})d^3r }
\end{equation}
should vanish (while in practice we demand it to be less than $0.1\ \mathrm{keV}^2$), so that Eq.~(\ref{equ-13}) is fulfilled. Note that in our calculation the global charge neutrality condition are fulfilled, i.e.,
\begin{equation}\label{equ-19}
   \int[n_p(\Vec{r})-n_e(\Vec{r})-n_\mu(\Vec{r})]d^3r \equiv 0.
\end{equation}
In addition, we have demanded that the $\beta$-equilibrium condition for the leptons is always fulfilled, i.e.,
\begin{equation}\label{equ-19}
   \mu_e(\Vec{r})=\mu_\mu(\Vec{r}).
\end{equation}
Equations~(\ref{equ-15})-(\ref{equ-17}) usually set five types of geometrically symmetric pasta phases assuming different dimensions, i.e. 1D for slab phase, 2D for the rod-tube phases, and 3D for the droplet-bubble phases. Once the density profiles are fixed, the droplet size $R_\mathrm{d}$ and WS cell size $R_\mathrm{W}$ are determined by
\begin{equation}\label{equ-20}
   R_\mathrm{d}=\left\{
\begin{aligned}
 & R_\mathrm{W}u^{1/\text{D}}, \text{droplet-like} \\
 & R_\mathrm{W}\left(1-u\right)^{1/\text{D}}, \text{bubble-like}
\end{aligned}
\right.,
\end{equation}
where the filling factor $u = {\langle n_p\rangle^2}/{\langle n_p^2\rangle}$ with $\langle n_p^2\rangle=\int n_p^2(\Vec{r})d^3r/V $ and $\langle n_p\rangle=\int n_p(\Vec{r})d^3r/V$. The WS cell volume $V$ is obtained with
\begin{equation}\label{equ-21}
    V=\left\{
\begin{aligned}
 & \frac{4}{3}\pi R^3_\mathrm{W},\ \ \ \text{D}=3 \\
 &  \pi a R^2_\mathrm{W},\ \ \ \text{D}=2 \\
 &  a^2 R_\mathrm{W},\ \ \ \ \text{D}=1
\end{aligned}
\right..
\end{equation}
Here we have adopted a finite cell size $a=30$ fm at dimensions $\text{D}=1$ and $2$ so that the volumes of slabs and rods/tubes are finite. Among the five types of nonuniform structures with various WS cell sizes $R_\mathrm{W}$ and the uniform phase for dense stellar matter, the optimal structure is obtained by searching for the configuration that minimizes the energy per baryon $E/A$ at fixed $Y_p$ and $n_\text{b}$.

Finally, it should be mentioned that at densities $n_\text{b}<10^{-4}\ \mathrm{fm^{-3}}$ the WS cell size $R_\mathrm{W}$ becomes too large so that viable numerical simulations become difficult. In such cases, we divide the WS cell of the droplet phase into two parts, i.e., a core ($r\leq 25.6$ fm) enclosing a nucleus with nonuniform density distributions and a spherical shell ($25.6\ \mathrm{fm}<r\leq R_\mathrm{W}$) comprised of neutrons, protons, and electrons at constant densities, while the density profiles still follow the constancy of chemical potentials indicated by Eq.~(\ref{equ-13}) with the chemical potentials in the shell region take their average values. More detailed discussion on the treatment can be found in Ref.~\cite{Xia2022_PRC105-045803}.

\section{\label{sec:res}Results and Discussion}

\begin{figure}
    \includegraphics[width=0.8\linewidth]{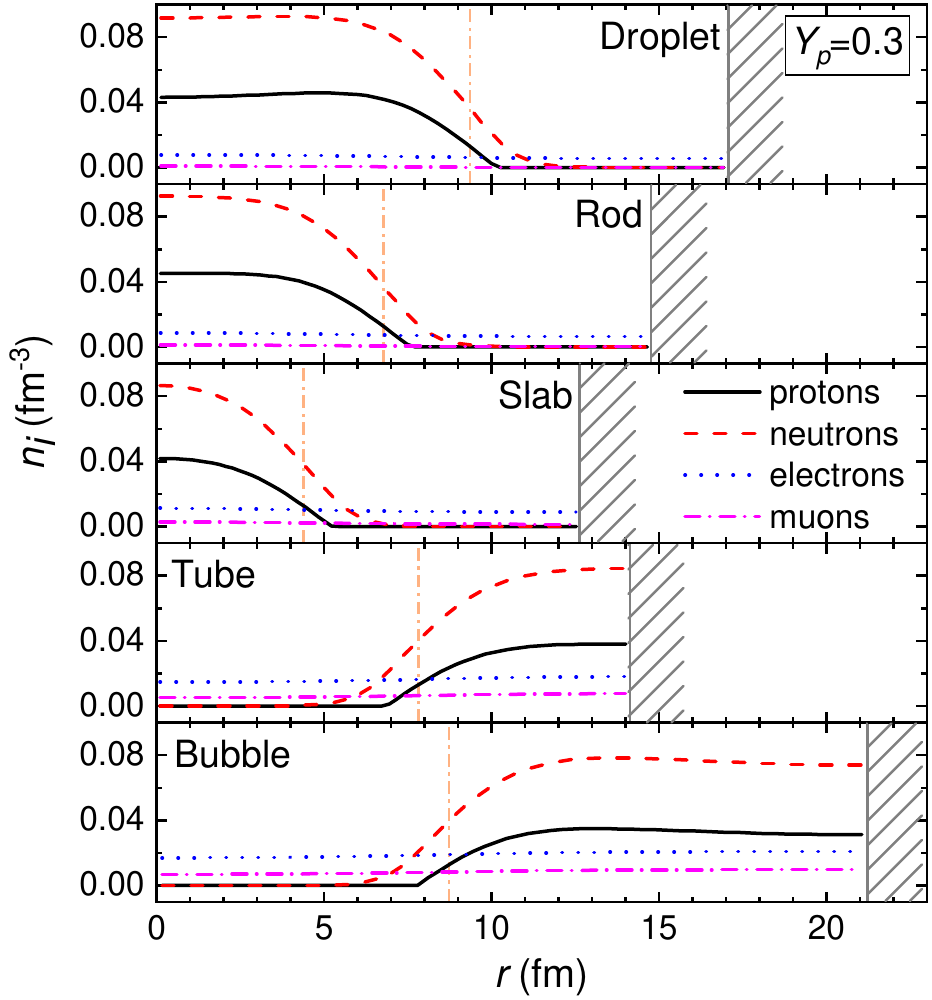}
    \caption{\label{fig:rholine}Density profiles of protons, neutrons, electrons, and muons for the droplet, rod, slab, tube, and bubble phases at $\text{log}_{10}(n_\mathrm{b}/\mathrm{fm^{-3}})={-1.7}, {-1.6}, {-1.4}, {-1.1}, {-1}$ and $Y_p=0.3$, where the RDF DD-LZ1 is employed~\cite{Wei2020_CPC44-074107}. The boundaries of the WS cells $R_\mathrm{W}$ are indicated by shaded region and the vertical orange dotted lines represent the droplet size $R_\mathrm{d}$ fixed by Eq.~(\ref{equ-20}).}
\end{figure}

Based on the formalism introduced in Sec.~\ref{sec:Theor}, we can then examine the properties and microscopic structures of dense stellar matter at given densities $n_\text{b}$ and proton fractions $Y_p$. As an example, in Fig.~\ref{fig:rholine} we present the typical density profiles for the droplet, rod, slab, tube, and bubble phases inside Wigner-Seitz cells at fixed proton fraction $Y_p=0.3$, where the RDF DD-LZ1 is adopted~\cite{Wei2020_CPC44-074107}. As density increases, the droplet, rod, slab, tube, and bubble phases appear sequentially. The proton and neutron densities in the droplet/rod phases are larger at $r=0$ than those at $r=R_\mathrm{W}$, while that of the tube/bubble phases are the opposite. Based on the density profiles, the droplet size $R_\mathrm{d}$ can then be fixed by Eq.~(\ref{equ-20}), which are indicated by the vertical orange dotted lines in Fig.~\ref{fig:rholine}. Since the masses of electrons and muons are small with the Compton wavelengths $\lambda_{e,\mu}> R_\mathrm{W}$, the density distributions of the leptons in WS cells are almost homogeneous, while the slight variations in densities are dominated by the Coulomb potential $A(r)$.

\begin{table}
\caption{\label{table:NM} Saturation properties of nuclear matter corresponding to the RDFs indicated in Table~\ref{table:param}. }
\begin{tabular}{c|ccccccc} \hline \hline
       & $n_0$        &   $B$    &   $K$  &  $J$   & $S$    &  $L$  & $K_\mathrm{sym}$        \\
       & fm${}^{-3}$  &   MeV    &   MeV  &  MeV   &  MeV   &  MeV  &   MeV             \\ \hline
NL3    &  0.148       & $-$16.25 &  271.7 &  204   & 37.4   & 118.6 &   101           \\
PK1    &  0.148       & $-$16.27 &  282.7 &$-27.8$ & 37.6   & 115.9 &   55           \\
PK1r   &  0.148       & $-$16.27 &  283.7 &$-19.1$ & 37.8   & 116.5 &   56           \\
GM1    &  0.153       & $-$16.33 &  300.5 & $-216$ & 32.5   &  94.0 &   18           \\
MTVTC  &  0.153       & $-$16.30 &  239.8 & $-513$ & 32.5   &  89.6 &  $-6.5$         \\  \hline
PKDD   &  0.150       & $-$16.27 &  262.2 & $-119$ & 36.8   &  90.2 &  $-81$           \\
DD-MEX2&  0.152       & $-$16.09 &  255.1 & 610    & 35.3   &  87.0 &  $-55$           \\
TW99   &  0.153       & $-$16.24 &  240.2 & $-540$ & 32.8   &  55.3 &  $-125$          \\
DD-MEY &  0.153       & $-$16.11 &  264.3 & 745    & 32.2   &  53.8 &  $-78$           \\
DD-MEX1&  0.151       & $-$16.03 &  292.1 & 941    & 31.8   &  53.3 &  $-66$           \\
DD-ME2 &  0.152       & $-$16.13 &  250.8 & 477    & 32.3   &  51.2 &  $-87$           \\
DD-MEX &  0.152       & $-$16.06 &  266.6 & 875    & 32.2   &  49.6 &  $-72$           \\
DD-LZ1 &  0.158       & $-$16.06 &  230.7 & 1330   & 32.0   &  42.5 &  $-20$            \\
\hline
\end{tabular}
\end{table}

In this work, as indicated in Table~\ref{table:param}, we adopt in total 13 different RDFs including those with nonlinear self-couplings (NL3~\cite{Lalazissis1997_PRC55-540}, PK1~\cite{Long2004_PRC69-034319},  PK1r~\cite{Long2004_PRC69-034319}, GM1~\cite{Glendenning1991_PRL67-2414}, MTVTC~\cite{Maruyama2005_PRC72-015802}) and density-dependent couplings (DD-LZ1~\cite{Wei2020_CPC44-074107}, PKDD~\cite{Long2004_PRC69-034319}, DD-ME2~\cite{Lalazissis2005_PRC71-024312}, TW99~\cite{Typel1999_NPA656-331}, DD-MEX~\cite{Taninah2020_PLB800-135065}, DD-MEX1, DD-MEX2, and DD-MEY~\cite{Taninah2023_PRC107-041301}). Neglecting the contributions of leptons and Coulomb interactions, the binding energy per nucleon $\varepsilon\left(n_\mathrm{b}, Y_p\right)$ of uniform nuclear matter can be expanded as
\begin{eqnarray}
\varepsilon &=& \varepsilon_0\left(n_\mathrm{b}\right) + \varepsilon_\mathrm{sym}\left(n_\mathrm{b}\right) \left(1-2Y_p\right)^2, \\
\varepsilon_0 &=&B+\frac{K}{18} \left ( \frac{n_\mathrm{b}}{n_0}-1  \right ) ^{2}+\frac{J}{162} \left ( \frac{n_\mathrm{b}}{n_0}-1  \right )^{3}, \label{eq:e0} \\
\varepsilon_\mathrm{sym} &=&S+\frac{L}{3} \left ( \frac{n_\mathrm{b}}{n_0}-1  \right )+\frac{K_{\mathrm{sym}}}{18} \left ( \frac{n_\mathrm{b}}{n_0}-1  \right )^{2}.
\label{eq:esym}
\end{eqnarray}
Here $n_0$ represents the nuclear saturation density, $B$ the binding energy, $K$ the incompressibility, $J$ the skewness, $S$ the symmetry energy, $L$ and $K_\mathrm{sym}$ the slope and curvature of symmetry energy. The corresponding values for the saturation properties of nuclear matter predicted by the 13 RDFs are indicated in Table.~\ref{table:NM}, which generally cover the current uncertainty bands~\cite{Shlomo2006_EPJA30-23, Li2013_PLB727-276, Oertel2017_RMP89-015007, PREX2021_PRL126-172502, CREX2022_PRL129-042501, Zhang2020_PRC101-034303, Essick2021_PRL127-192701, Reinhard2022_PRL129-232501, Danielewicz2002_Science298-1592, Liu2021_PRC103-014616, Xie2021_JPG48-025110}.

\subsection{\label{sec:res_2D}Cold dense stellar matter}

\begin{table}
    \caption{\label{tab:number}Data point numbers and ranges of $n_\text{b}$ and $Y_p$ for each EOS data table (attached files ``{RDF\_2D.dat}") of cold dense stellar matter predicted by various RDFs. }
    \begin{tabular}{c|ccc}
    \hline\hline
     Parameter & Range & Step & Total \\
     \hline
     $\text{log}_{10}(n_\mathrm{b}/\mathrm{fm^{-3}})$& $-8$ to $-0.8$& 0.1 &  73 \\
      $n_\mathrm{b}/\mathrm{fm^{-3}}$ & 0.16 to 2  & 0.002 &  921 \\
     \hline
     $Y_p$   & 0.01 to 0.65  & 0.01 &  65 \\
     \hline
    \end{tabular}
\end{table}

We then carry out systematic investigations on the properties and microscopic structures of dense stellar matter at various densities $n_\text{b}$ and proton fractions $Y_p$. In particular, as indicated in Table \ref{tab:number}, the following partitioning of parameter space for the EOSs are employed in our calculation, i.e.,
\begin{enumerate}
  \item Proton fraction $Y_p$: A step of $0.01$ ranging from $Y_p=0.01$ to $0.65$ is used, resulting in total $65$ grid points for $Y_p$.
  \item Baryon number density $n_\text{b}$: A step of $0.01$ ranging from $\text{log}_{10}(n_\text{b}/\mathrm{fm}^{-3})=-8$ to $-0.8$ and at larger densities a step of 0.002 $\mathrm{fm}^{-3}$ ranging from $n_\text{b}=0.16\ \mathrm{fm}^{-3}$ to 2 $\mathrm{fm}^{-3}$ are adopted, resulting in total $994$ grid points for $n_\text{b}$.
\end{enumerate}
Then there are in total 64610 data points of dense stellar matter calculated for each RDF, where various physical properties such as the energy per baryon $E/A$, pressure $P$, chemical potentials $\mu_i$, droplet radius $R_\mathrm{d}$, WS cell size $R_\mathrm{W}$, proton number $N_p$, densities of neutrons and protons $n_{n, p} (r)$ at $r=0$ and $R_\mathrm{W}$  are obtained for fixed proton fraction $Y_p$ and baryon number density $n_\text{b}$.

\subsubsection{\label{sec:res_2D_DDLZ1}Predictions from the RDF DD-LZ1}

\begin{figure}
    \includegraphics[width=\linewidth]{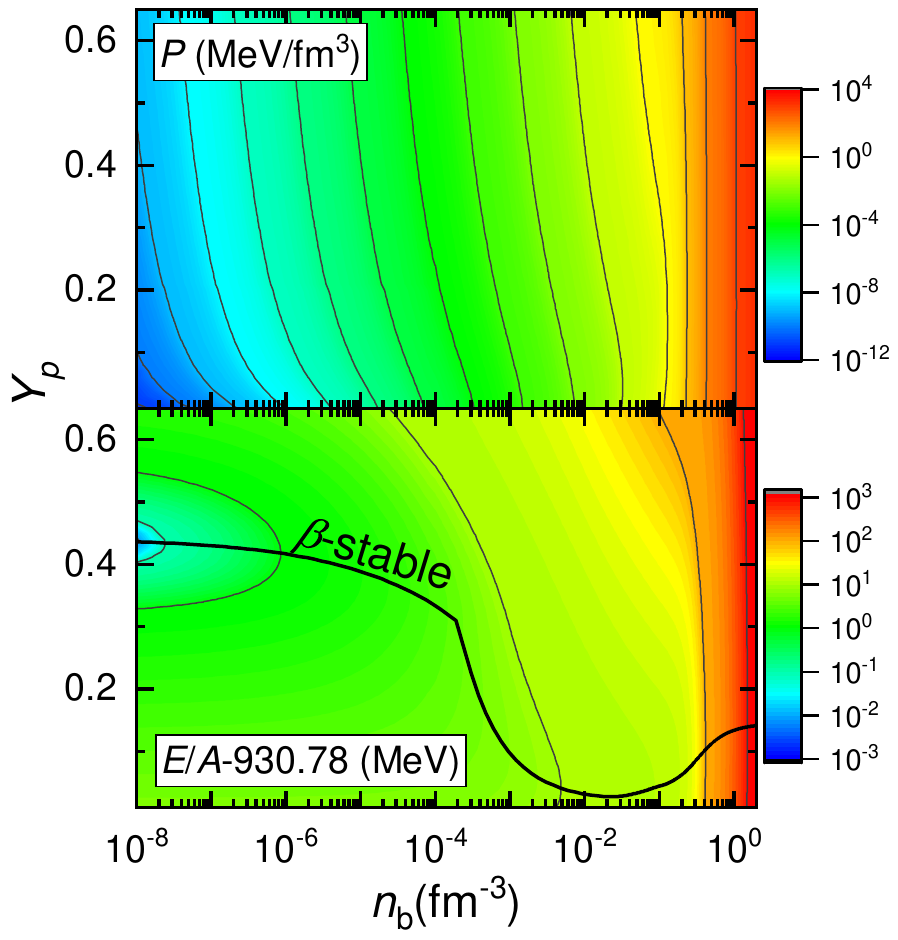}
    \caption{\label{fig:EOS_DDLZ1} Pressure $P$ and energy per baryon $E/A$ for cold dense stellar matter predicted by the RDF DD-LZ1~\cite{Wei2020_CPC44-074107}. The $\beta$-stability line fixed by $\mu_n=\mu_p + \mu_e$ is indicated by the black solid curve.}
\end{figure}

To illustrate the obtained results, in Fig.~\ref{fig:EOS_DDLZ1} we present the pressure $P$ and energy per baryon $E/A$ of cold dense stellar matter as functions of baryon number density $n_\text{b}$ and proton fraction $Y_p$, which are fixed by adopting the RDF DD-LZ1~\cite{Wei2020_CPC44-074107}. It is found that the pressure and energy per baryon of dense stellar matter are generally increasing with density, where $P$ ranges 16 orders of magnitude from $10^{-12} \ \mathrm{MeV/fm}^{3}$ to $10^{4} \ \mathrm{MeV/fm}^{3}$ as we increase $n_\text{b}$ from $10^{-8} \ \mathrm{fm}^{-3}$ to 2 $\mathrm{fm}^{-3}$. At small densities with $n_\text{b}\lesssim 10^{-3} \ \mathrm{fm}^{-3}$, the pressure is dominated by the degenerate electron gas, so that $P$ generally increases with $Y_p$. At densities $n_\text{b}\gtrsim 0.2 \ \mathrm{fm}^{-3}$, the dense stellar matter becomes uniform and the pressure is dominated by nuclear matter, which varies little with $Y_p$. Nevertheless, the variation of the energy per baryon $E/A$ with respect to $Y_p$ is nonmonotonic. In particular, as indicated by the solid curve in Fig.~\ref{fig:EOS_DDLZ1}, a minimum value for $E/A$ at fixed $n_\text{b}$ can be obtained fulfilling the $\beta$-stability condition $\mu_n=\mu_p + \mu_e$, which corresponds to neutron star matter and will be examined in Sec.~\ref{sec:res_1D}. At vanishing densities, neutron star matter is comprised of ions emersed in an electron gas, where the corresponding minimum energy per baryon is $\sim$930.6 MeV. Note that there exist a kink for the $\beta$-stability line at $n_\text{b}\approx 10^{-4} \ \mathrm{fm}^{-3}$, which corresponds to the neutron drip density with neutron gas emerges outside of nuclei. The variation of energy per baryon is relatively small at $n_\text{b}\lesssim 10^{-3} \ \mathrm{fm}^{-3}$ and resulting in small pressures with $P=n_\text{b}^2 \mbox{d} (E/A)/\mbox{d} n_\text{b}$, which becomes significant at larger densities.

\begin{figure}
    \includegraphics[width=0.9\linewidth]{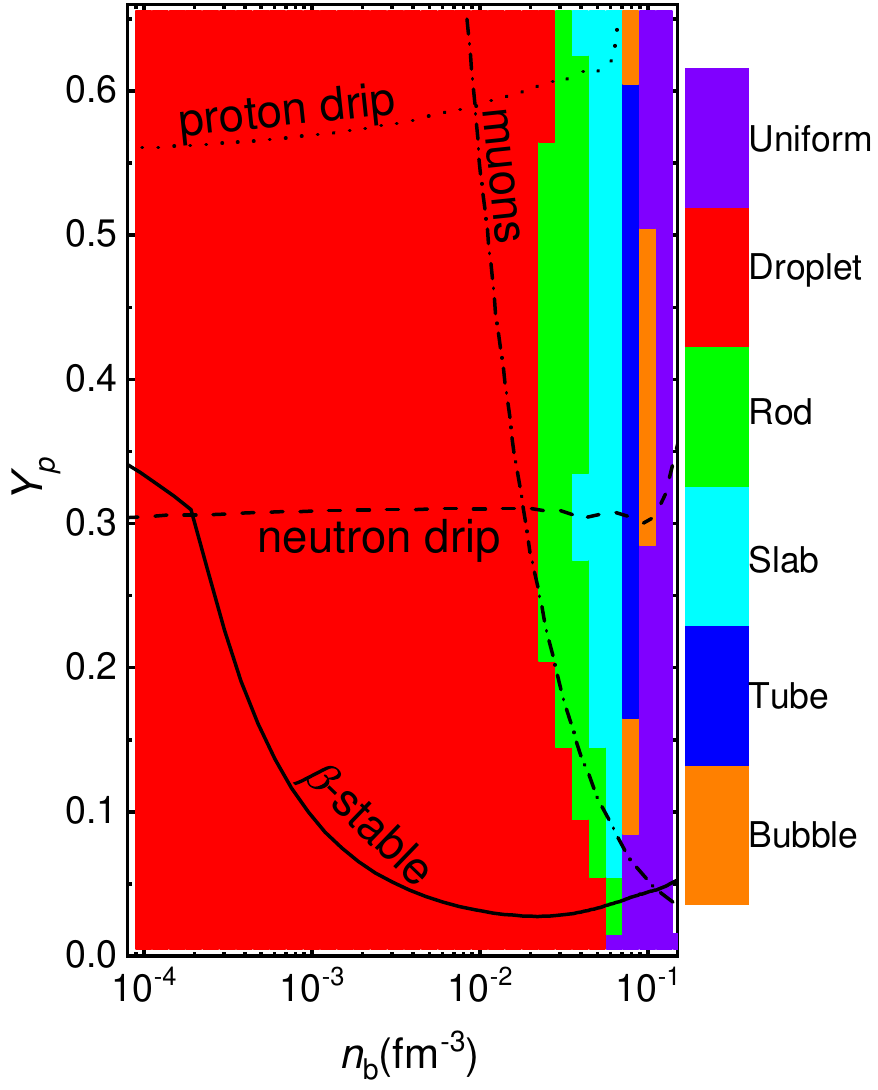}
    \caption{\label{fig:dimn} Phase diagram for cold dense stellar matter predicted by the RDF DD-LZ1~\cite{Wei2020_CPC44-074107}, where the corresponding EOS is indicated in Fig.~\ref{fig:EOS_DDLZ1}. The $\beta$-stability line is indicated by the black solid curve. The dashed (dotted) curve indicates the critical condition for neutron (proton) drip with $\mu_{n}\geq m_{n}$ ($\mu_{p}\geq m_{p}$), where neutron (proton) gas outside of nuclei emerges at smaller (larger) $Y_p$. The onset densities of muons with $\mu_e \geq m_\mu$ are also indicated by the dash-dotted curve, above which muons emerge.}
\end{figure}

In Fig.~\ref{fig:dimn} we present the phase diagram of cold dense stellar matter predicted by the RDF DD-LZ1~\cite{Wei2020_CPC44-074107}. At small densities with $n_\text{b}\lesssim0.025 \ \mathrm{fm}^{-3}$, dense stellar matter generally exhibits droplet phase, while at larger densities the rod, slab, tube, and bubble phases emerge sequentially as they become energetically more favorable. Finally, at densities larger than $\sim$$0.1 \ \mathrm{fm}^{-3}$, nonuniform structures become unstable and dense stellar matter turns into uniform phase. The density ranges of various types of nonuniform structures vary with $Y_p$, which is largest at $Y_p\approx 0.3$ and decreases at $Y_p\lesssim 0.2$ or $Y_p\gtrsim 0.5$.

\begin{figure}
    \includegraphics[width=\linewidth]{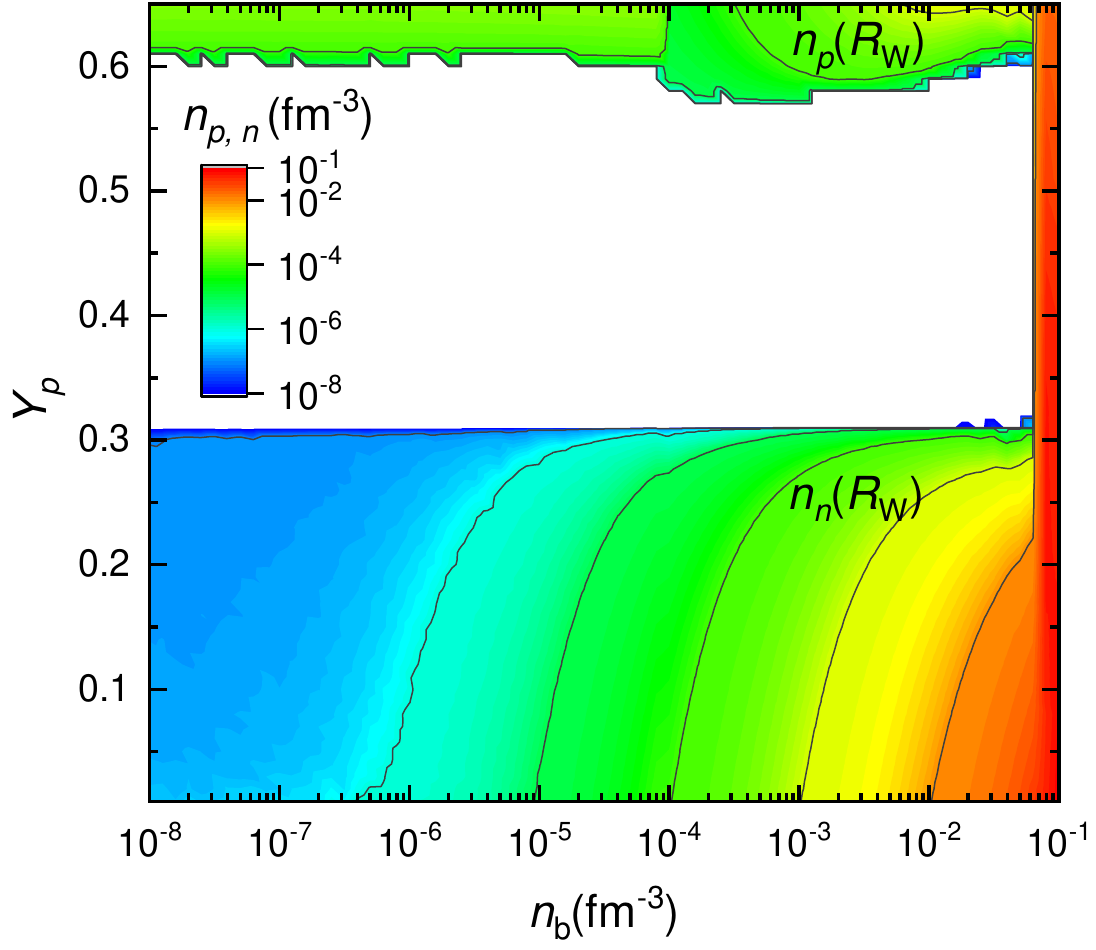}
    \caption{\label{fig:nout_DDLZ1} Densities of neutrons $n_n(R_\mathrm{W})$ and protons $n_p(R_\mathrm{W})$ at the boundary of WS cells, where the corresponding neutron and proton drip densities are indicated by the dashed and dotted curves in Fig.~\ref{fig:dimn}.}
\end{figure}

Beside the variations of microscopic structures in dense stellar matter with respect to $n_\text{b}$ and $Y_p$, the emergence of new degrees of freedom will also take place. For example, as indicated by the dashed (dotted) curve in Fig.~\ref{fig:dimn}, by examine the chemical potentials of neutrons (protons), the neutron (proton) gas outside of nuclei appears once $\mu_{n}\geq m_{n}$ ($\mu_{p}\geq m_{p}$). The corresponding densities of neutron and proton gases at the boundaries of WS cells are indicated in Fig.~\ref{fig:nout_DDLZ1}, which emerge at $Y_p\lesssim 0.3$ for neutron gas and $Y_p\gtrsim 0.56$ for proton gas. It is found that neutron and proton densities  $n_n(R_\mathrm{W})$ and $n_p(R_\mathrm{W})$ are generally increasing with $n_\text{b}$. Note that there are slight discontinuities for $n_n(R_\mathrm{W})$ and $n_p(R_\mathrm{W})$ at $n_\text{b}=10^{-4} \ \mathrm{fm}^{-3}$, which are caused by dividing a WS cell into two parts. Additionally, as indicated by the dash-dotted curve in Fig.~\ref{fig:dimn}, muons will inevitable appear with $\mu_e \geq m_\mu$ as we increase the density. The emergence of the new degrees of freedom as well as the variations of microscopic structures in dense stellar matter are expected to affect various astrophysical processes in core-collapse supernova, neutron stars, and binary neutron star mergers, e.g., the $g$-mode oscillations in neutron stars~\cite{Sun2025_PRD111-103019}.

\begin{figure}
    \includegraphics[width=\linewidth]{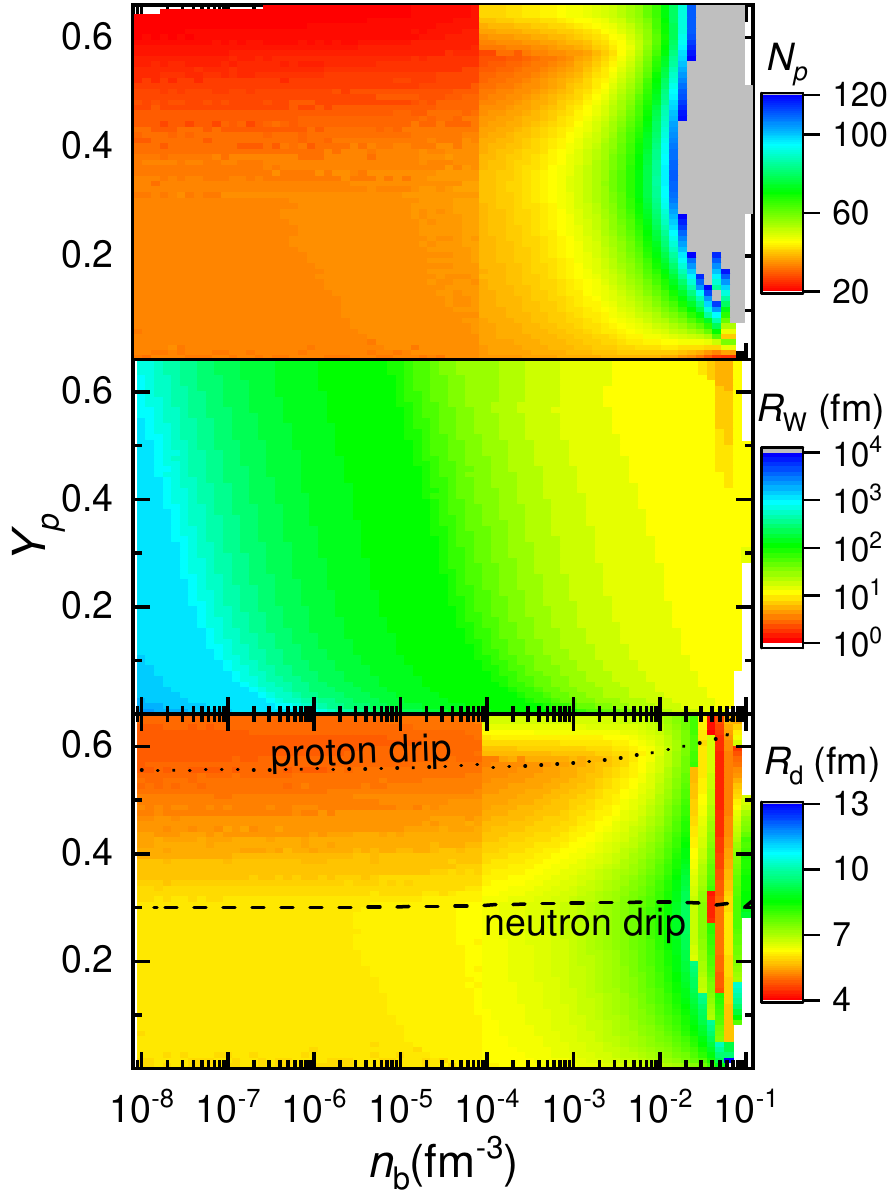}
    \caption{\label{fig:ZRdW_DDLZ1} Proton number $N_p$, droplet radius $R_\mathrm{d}$, and WS cell size $R_\mathrm{W}$ for the nonuniform structures in cold dense stellar matter, where the corresponding EOS is indicated in Fig.~\ref{fig:EOS_DDLZ1}.}
\end{figure}

The variations of microscopic structures in dense stellar matter are expected to alter the transport and elastic properties for various astrophysical processes~\cite{Horowitz2016, Chamel2008_LRR11-10, Caplan2017_RMP89-041002}, where in Fig.~\ref{fig:ZRdW_DDLZ1} we present the obtained proton number $N_p$, WS cell radius $R_\mathrm{W}$, and droplet size $R_\mathrm{d}$ predicted by the RDF DD-LZ1~\cite{Wei2020_CPC44-074107}. The proton numbers $N_p$ and droplet sizes $R_\mathrm{d}$ for nuclei in droplet phase share similar trends, which generally increase with density $n_\mathrm{b}$ and reach their maximum values at $Y_p\approx 0.3$ around the neutron drip line. In particular, for the droplet phase at $n_\mathrm{b} \lesssim 10^{-4}$ fm${}^{-3}$, the obtained proton numbers are generally $N_p\approx 33$ at $Y_p \lesssim 0.3$ and quickly drop to $\sim$20 as we increase $Y_p$, while $R_\mathrm{d}\approx 6.2$ fm at $Y_p \lesssim 0.3$ and quickly drop to $\sim$4.8 fm at $Y_p \approx 0.6$. Note that there exist slight discrepancies for $N_p$ and $R_\mathrm{d}$ at $n_\mathrm{b} < 10^{-4}$ fm${}^{-3}$, which become sizable once proton drip take place at $Y_p \gtrsim 0.6$. This is mainly caused by dividing the WS cell of the droplet phase into two parts~\cite{Xia2022_PRC105-045803}, which should be improved in our future study. The variations of $N_p$ and $R_\mathrm{d}$ become more significant at larger densities before the transition of droplet phase into rod phase takes place, where $N_p$ and $R_\mathrm{d}$ could reach $\sim$100 and $\sim$10 fm. Comparing with different nonuniform structures, the droplet sizes $R_\mathrm{d}$ decline for droplets/bubbles, rods/tubes, and slabs as the dimension D decreases. The WS cell size $R_\mathrm{W}$ is generally decreasing with $Y_p$ and $n_\mathrm{b}$, which reaches $10^4$ fm and is expected to grow at smaller $n_\mathrm{b}$ since the number of nucleons in WS cells is almost constant at fixed $Y_p$. Similar to $N_p$ and $R_\mathrm{d}$, slight discrepancies are also observed at $n_\mathrm{b} < 10^{-4}$ fm${}^{-3}$ once proton drip takes place at $Y_p \gtrsim 0.6$.

\subsubsection{\label{sec:res_2D_All} Comparison with the other RDFs}

As indicated in Table~\ref{table:param}, the nuclear matter properties vary with the adopted RDF, where the higher order coefficients in Eqs.~(\ref{eq:e0}) and (\ref{eq:esym}) cover a large range, i.e., the incompressibility $K\approx 230$-300 MeV, the skewness $J\approx -500$-1300 MeV, the symmetry energy $S\approx 32$-38 MeV, the slope and curvature of symmetry energy $L\approx 42$-120 MeV and $K_\mathrm{sym}\approx -125$-100 MeV. The variations in nuclear matter properties are expected to alter the properties and microscopic structures of dense stellar matter~\cite{Xia2021_PRC103-055812, Xia2022_CTP74-095303, Xia2022_PRC105-045803}, which could affect various astrophysical processes.

\begin{figure*}
    \includegraphics[width=0.7\linewidth]{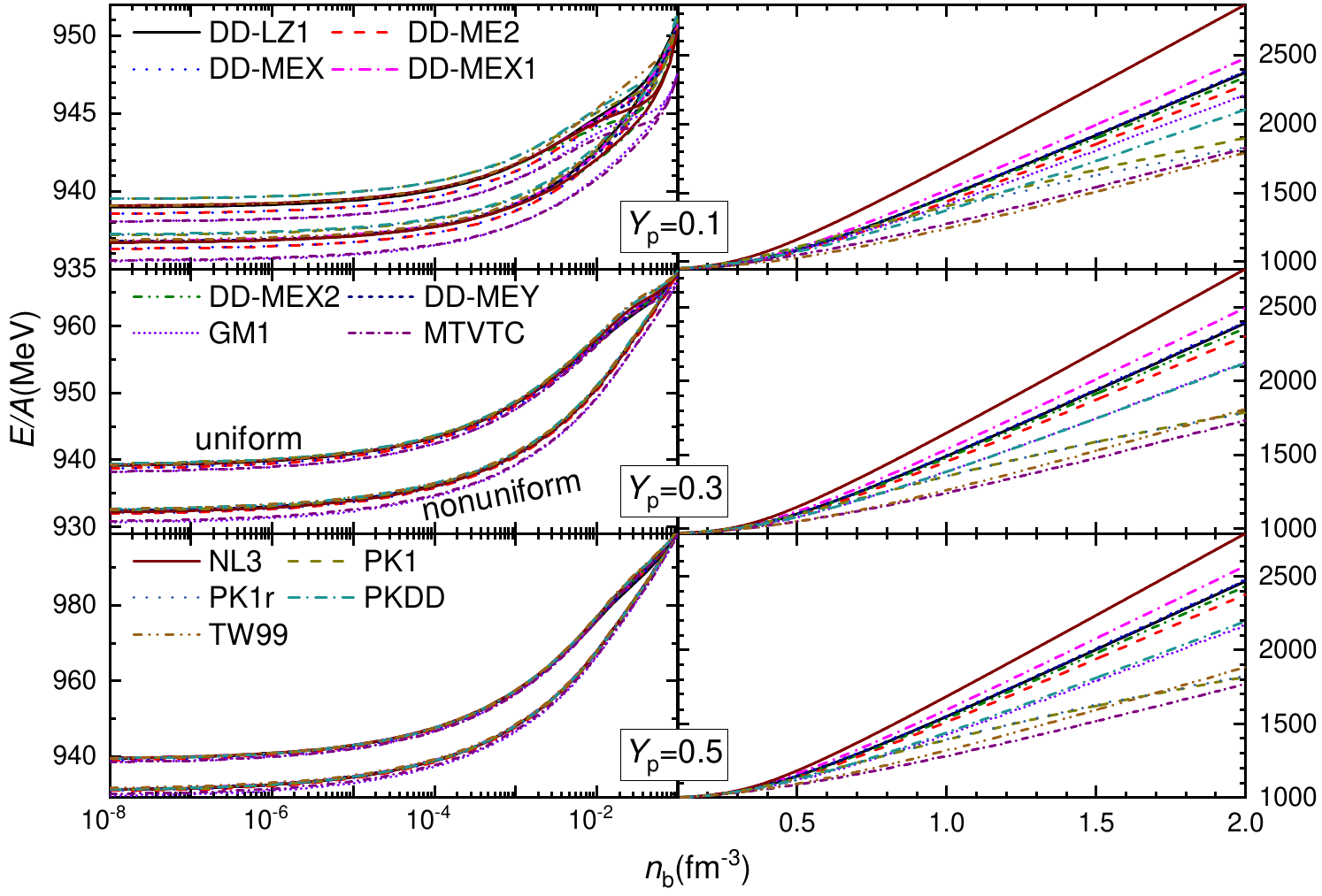}
    \caption{\label{fig:EpA} Energy per baryon $E/A$ between uniform and nonuniform matter as functions of baryon number density $n_\text{b}$ with $Y_p=0.1$, 0.3, and 0.5, where various RDFs indicated in Table~\ref{table:param} are adopted.}
\end{figure*}

In Fig.~\ref{fig:EpA} we present the obtained energy per baryon $E/A$ of dense stellar matter with fixed proton fractions $Y_p=0.1$, 0.3, and 0.5, which correspond to the most favorable nuclear shapes with optimum WS cell sizes $R_\mathrm{W}$. For comparison, the energy per baryon for uniform matter is presented, where at small densities nonuniform structures are more stable with the energy per baryon reduced by up to $\sim$8 MeV. In general, the energy reduction increases with $Y_p$, which is about 2 MeV at $Y_p=0.1$, 7 MeV at $Y_p=0.3$, and 8 MeV at $Y_p=0.5$. At larger densities with $n_\text{b}\gtrsim0.1 \ \mathrm{fm}^{-3}$, eventually the energy differences vanish, where uniform matter becomes more stable as indicated in Fig.~\ref{fig:dimn}. The differences of $E/A$ predicted by various RDFs at densities $n_\text{b}\lesssim 10^{-4}\ \mathrm{fm}^{-3}$ are mainly attributed to the differences in nucleon masses as indicated in Table~\ref{table:param}, while the obtained binding energies generally coincide with each other. At larger densities, the impact on binding energies due to the uncertainties of RDFs is insignificant at $n_\text{b}\lesssim0.1 \ \mathrm{fm}^{-3}$ and $Y_p\gtrsim 0.3$, which start to take effect at smaller $Y_p$ and $10^{-5}\ \mathrm{fm}^{-3} \lesssim n_\text{b}\lesssim0.1 \ \mathrm{fm}^{-3}$ fulfilling the critical condition for neutron drip with neutron gas formed outside of nuclei. If we further increase the density, as illustrated in the right panels in Fig.~\ref{fig:EpA}, dense stellar matter becomes uniform and the uncertainties in $E/A$ due to variations of RDFs grow quickly and become sizable.

\begin{figure*}
    \includegraphics[width=0.7\linewidth]{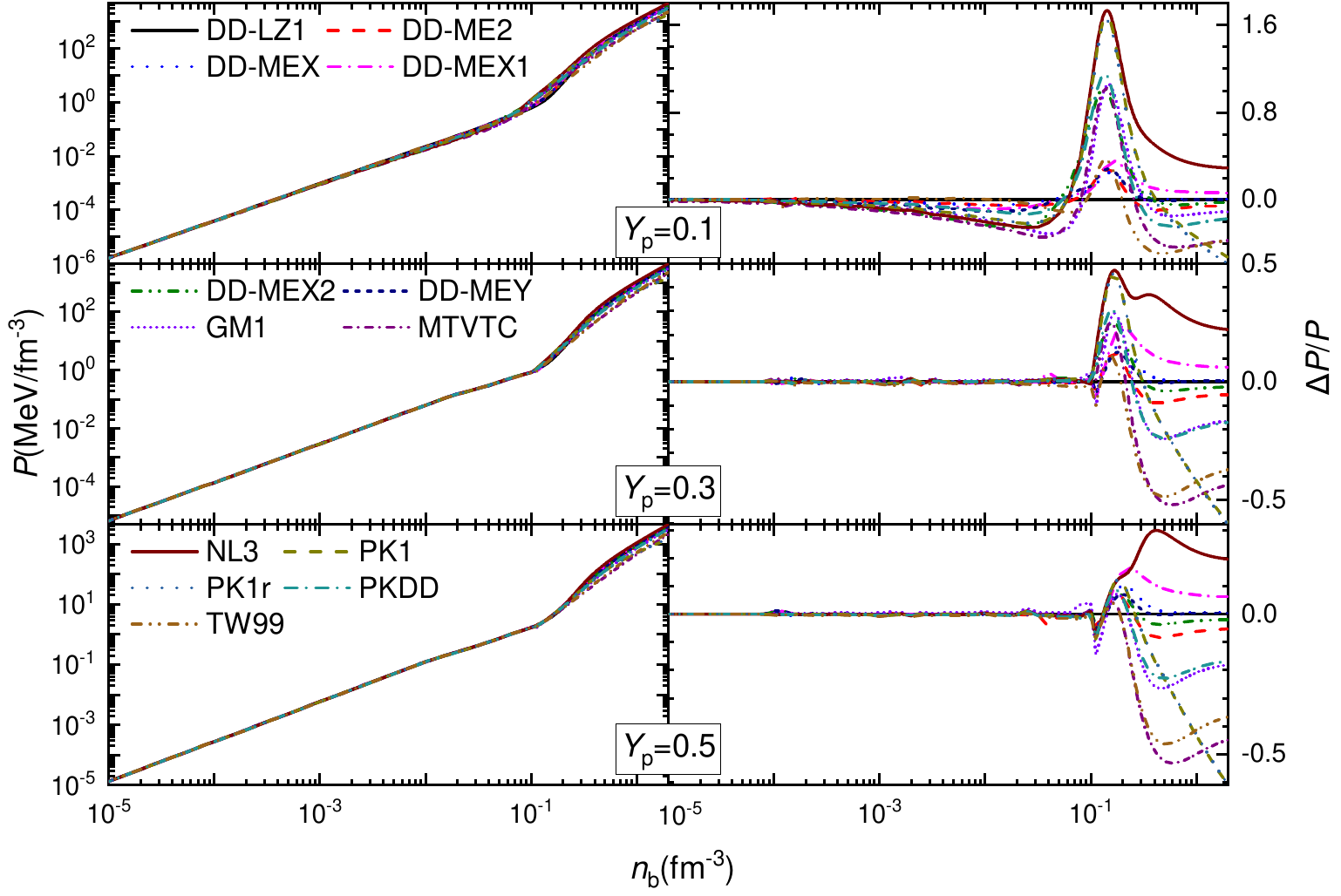}
\caption{\label{fig:Pressure} Similar as Fig.~\ref{fig:EpA} but for pressure $P$ (left) and its relative deviation $\Delta P/P$ from that of DD-LZ1 (right) with $\Delta P=P_{\text{DD-LZ1}}-P_{\mathrm{RDF}}$.}
\end{figure*}

In correspondence to Fig.~\ref{fig:EpA}, the pressure of dense stellar matter as functions of baryon number density $n_\text{b}$ with $Y_p=0.1$, 0.3, and 0.5 are presented in the left panels of Fig.~\ref{fig:Pressure}. To show the variations caused by the uncertainties of RDFs, in the right panels of Fig.~\ref{fig:Pressure} we have plotted the relative deviations of pressures $\Delta P/P$ from that of DD-LZ1, where $\Delta P=P_{\text{DD-LZ1}}-P_{\mathrm{RDF}}$ with $P_{\mathrm{RDF}}$ the pressure fixed by adopting one of the RDFs indicated in Table~\ref{table:param}. At vanishing densities with $n_\text{b}\lesssim 10^{-5}\ \mathrm{fm}^{-3}$, the dense stellar matter is comprised of Coulomb lattices of nuclei, electrons, and non-relativistic nucleons, the pressure is then dominated by the degeneracy pressure of electrons and nucleon gas, which is thus insensitive to the adopted RDFs and increases with $Y_p$. This is still true up to $n_\text{b}\approx 0.02\ \mathrm{fm}^{-3}$ at $Y_p\gtrsim 0.3$, while at smaller $Y_p$ neutrons start to drip out and the pressure varies with the adopted RDF. For dense stellar matter at $Y_p\lesssim 0.3$, the uncertainty of pressure due to the variations of RDFs grows with density and reaches its peak at $n_\text{b}\approx 0.02\ \mathrm{fm}^{-3}$, which coincides with our previous finding on the uncertainties of EOSs for neutron star matter~\cite{Xia2022_CTP74-095303}. At larger densities, the dense stellar matter becomes uniform, where the uncertainty of pressure grows significantly and varies with the adopted RDFs.

\begin{figure*}
    \includegraphics[width=0.33\linewidth]{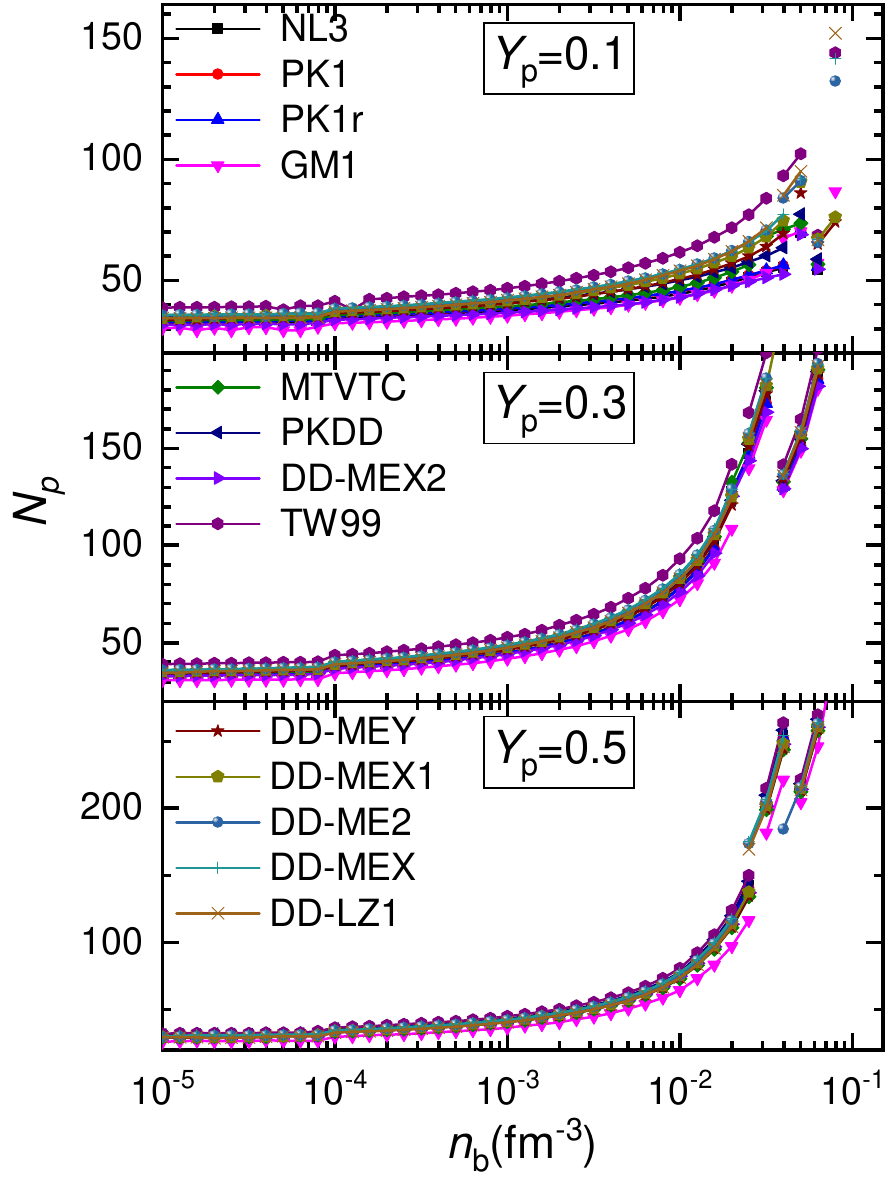}
    \includegraphics[width=0.323\linewidth]{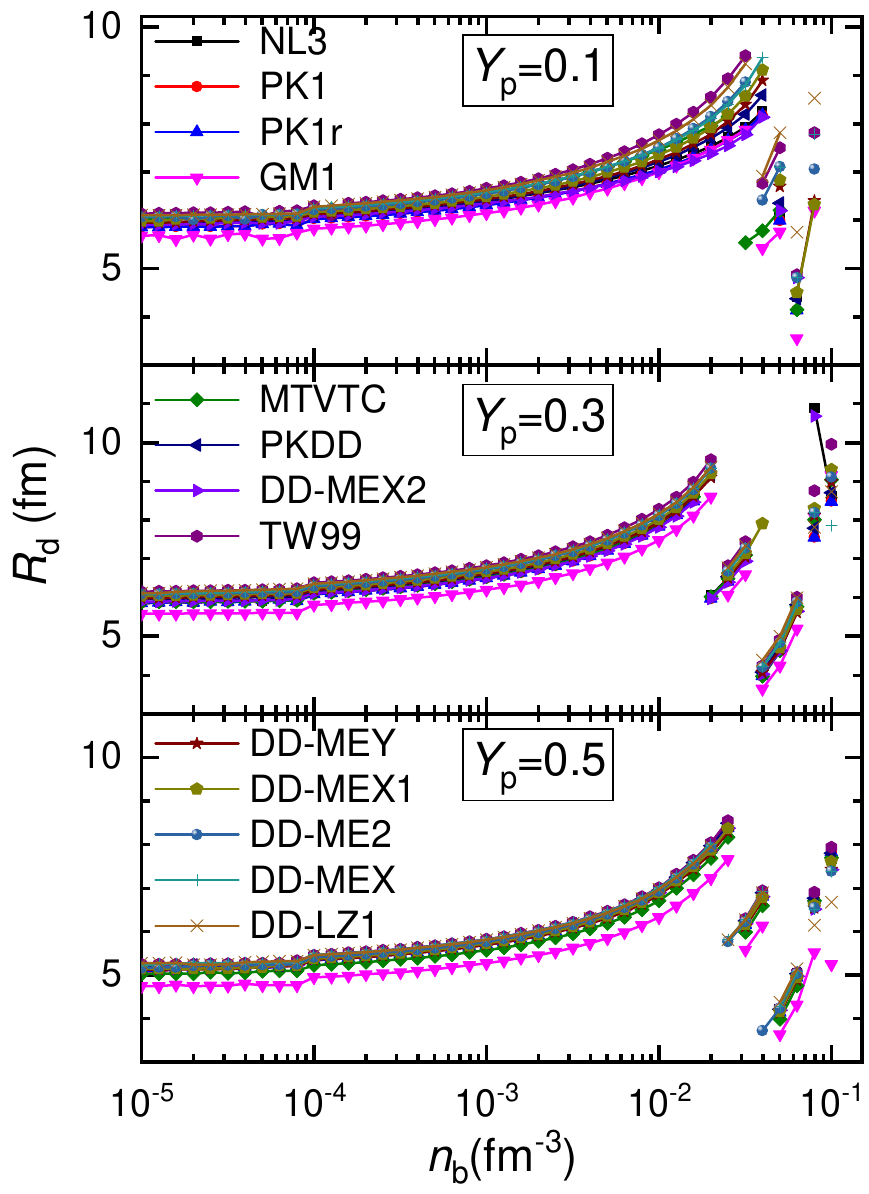}
    \includegraphics[width=0.33\linewidth]{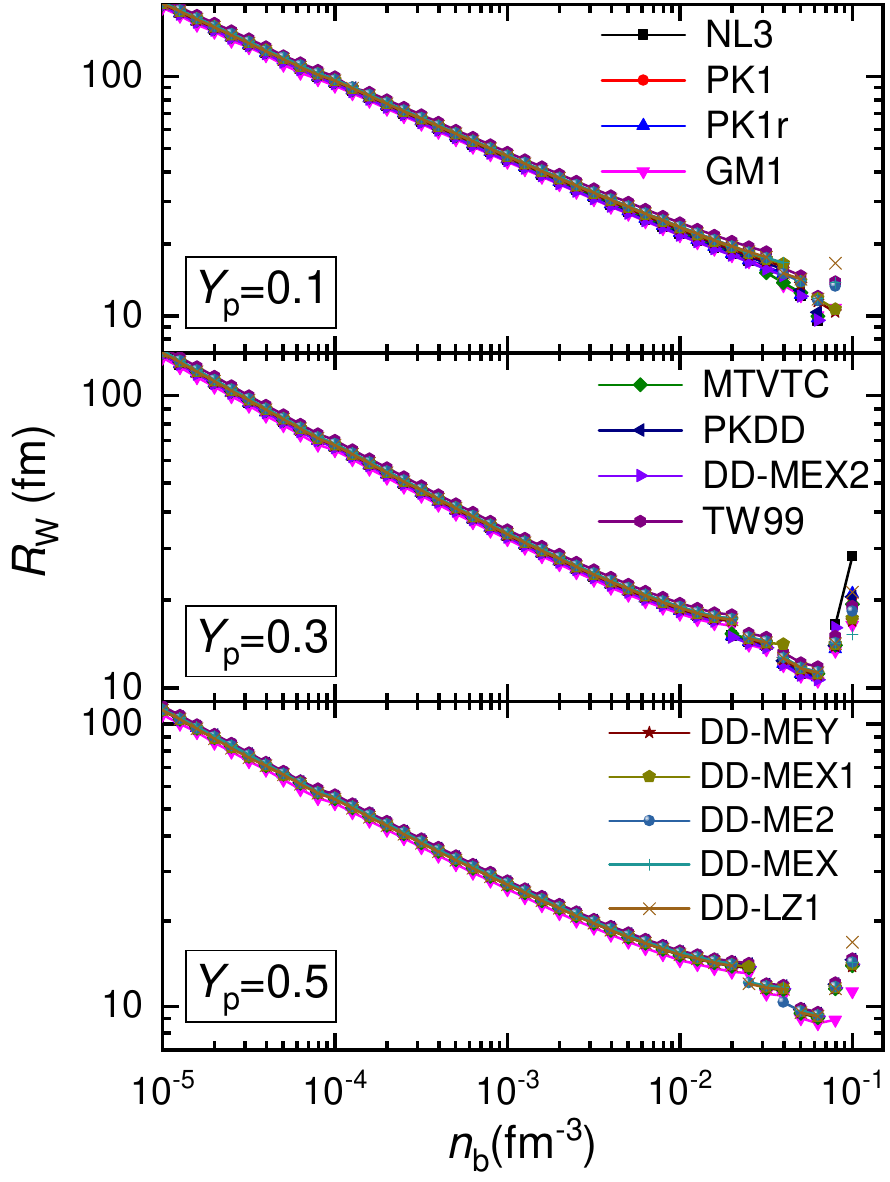}
\caption{\label{fig:Micro} Proton number $N_p$, droplet size $R_\mathrm{d}$, and WS cell radius $R_\mathrm{W}$ for the nonuniform structures in dense stellar matter, where the corresponding EOSs are indicated in Fig.~\ref{fig:EpA} and Fig.~\ref{fig:Pressure}.}
\end{figure*}

In Fig.~\ref{fig:Micro} we present the proton number $N_p$, droplet size $R_\mathrm{d}$, and WS cell radius $R_\mathrm{W}$ for the nonuniform structures in dense stellar matter. The variations of the microscopic structures with respect to $n_\mathrm{b}$ and $Y_p$ are generally similar to those predicted by DD-LZ1 as indicated in Fig.~\ref{fig:ZRdW_DDLZ1}. Similar to the EOSs of dense stellar matter, the variation of microscopic structures  with respect to the adopted RDFs are small at $n_\mathrm{b} \lesssim 10^{-4}$ fm${}^{-3}$. For example, slightly different proton numbers and droplet sizes are obtained at vanishing densities adopting various RDFs, which vary within the ranges $N_p\approx26$-40 and $R_\mathrm{d}\approx 4.8$-6.2 fm. The variation of WS cell radius $R_\mathrm{W}$ with respect to the adopted RDFs is generally limited within 10 \%  at $n_\mathrm{b} \lesssim 10^{-4}$ fm${}^{-3}$. At larger densities and smaller proton fractions ($Y_p\lesssim 0.3$) with the onset of neutron drip, the differences on the microscopic structures start to grow, where the values of $N_p$ and $R_\mathrm{d}$ as functions of density may exhibit different trends adopting various RDFs. Similar to previous investigations~\cite{Oyamatsu2007_PRC75-015801, Xu2009_ApJ697-1549, Grill2012_PRC85-055808, Bao2015_PRC91-015807, Shen2020_ApJ891-148, Xia2021_PRC103-055812}, the obtained values of $N_p$ and $R_\mathrm{d}$ approximately decrease with $L$ at $Y_p\lesssim 0.3$, while the values of $R_\mathrm{W}$ also decrease slightly. The nuclear shapes slightly vary with the adopted RDFs, where the variation become evident at $Y_p\lesssim 0.3$. In particular, more nuclear pasta structures emerge if RDFs with smaller $L$ are adopted at $Y_p\lesssim 0.3$. The density range of nuclear pasta also decreases with $L$ for dense stellar matter with $Y_p\lesssim 0.3$.

\begin{figure}
    \includegraphics[width=0.9\linewidth]{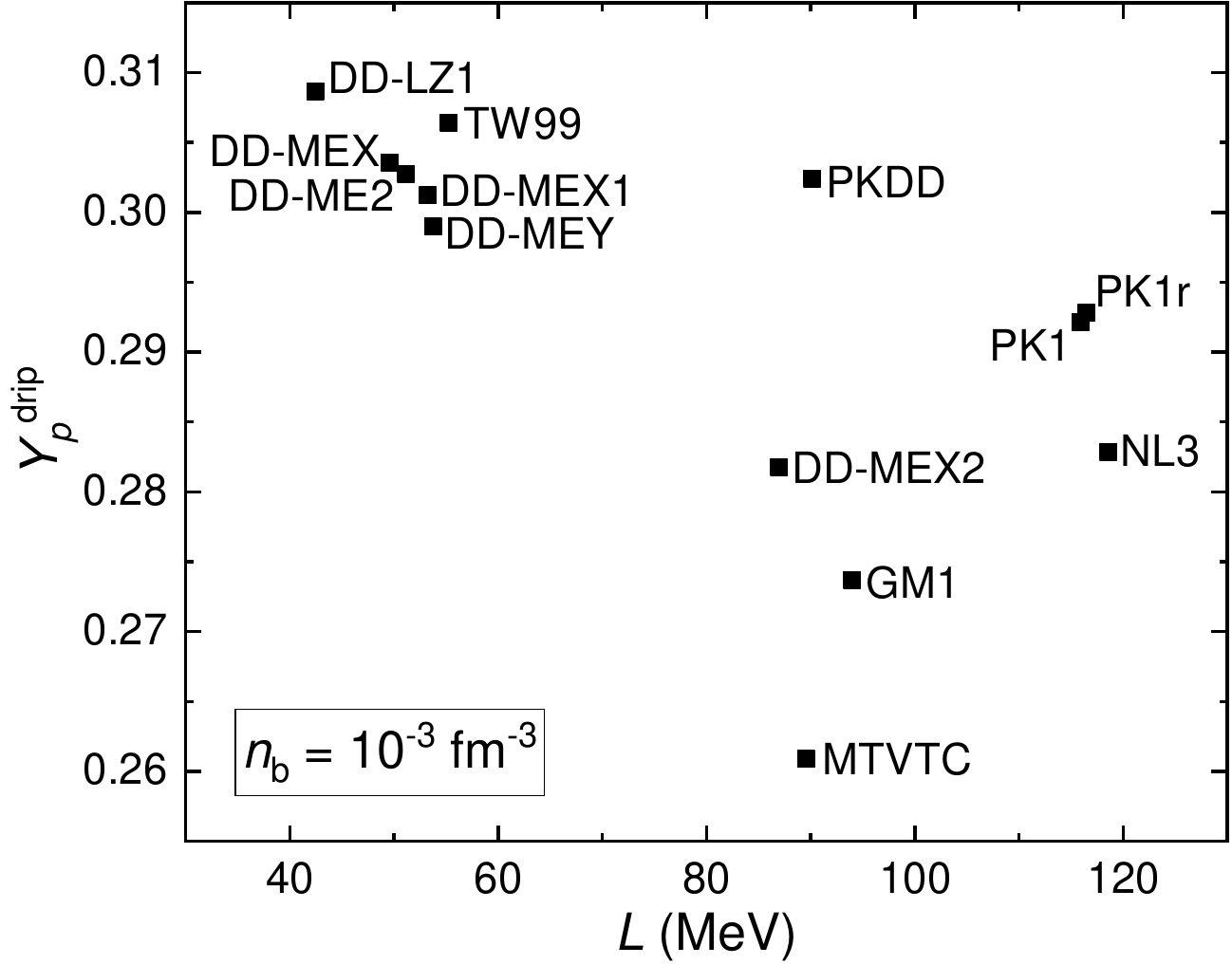}
    \caption{\label{fig:Ypdrip}The critical proton fractions $Y_p^\mathrm{drip}$ at $n_\text{b}=10^{-3} \ \mathrm{fm}^{-3}$ fixed by taking $\mu_{n}= m_{n}$, below which neutron drip starts to take place. }
\end{figure}

Finally, as indicated in Figs.~\ref{fig:EpA}-\ref{fig:Micro}, the EOSs and microscopic structures of dense stellar matter at $n_\text{b}\lesssim0.1 \ \mathrm{fm}^{-3}$ become sensitive to the adopted RDFs at $Y_p\lesssim 0.3$ with the onset of neutron drip. It is thus necessary to examine the critical condition for neutron drip. As indicated in Fig.~\ref{fig:dimn}, since the critical proton fraction of neutron drip $Y_p^\mathrm{drip}$ varies little with respect to the density at $n_\text{b}\lesssim0.1 \ \mathrm{fm}^{-3}$, we then fix the critical proton fractions predicted by various RDFs at $n_\text{b}=10^{-3} \ \mathrm{fm}^{-3}$. The obtained results are then presented in Fig.~\ref{fig:Ypdrip}, which are fixed by taking $\mu_{n}= m_{n}$ at $n_\text{b}=10^{-3} \ \mathrm{fm}^{-3}$ and neutron drip takes place at $Y_p<Y_p^\mathrm{drip}$. Evidently, the critical proton fractions lie within a small range with $0.26<Y_p^\mathrm{drip}<0.31$, which vary with the adopted RDFs. In general, the critical proton fraction $Y_p^\mathrm{drip}$ decreases with the slope of symmetry energy $L$.

\subsection{\label{sec:res_1D}Neutron star matter}

\begin{figure}
  \includegraphics[width=0.9\linewidth]{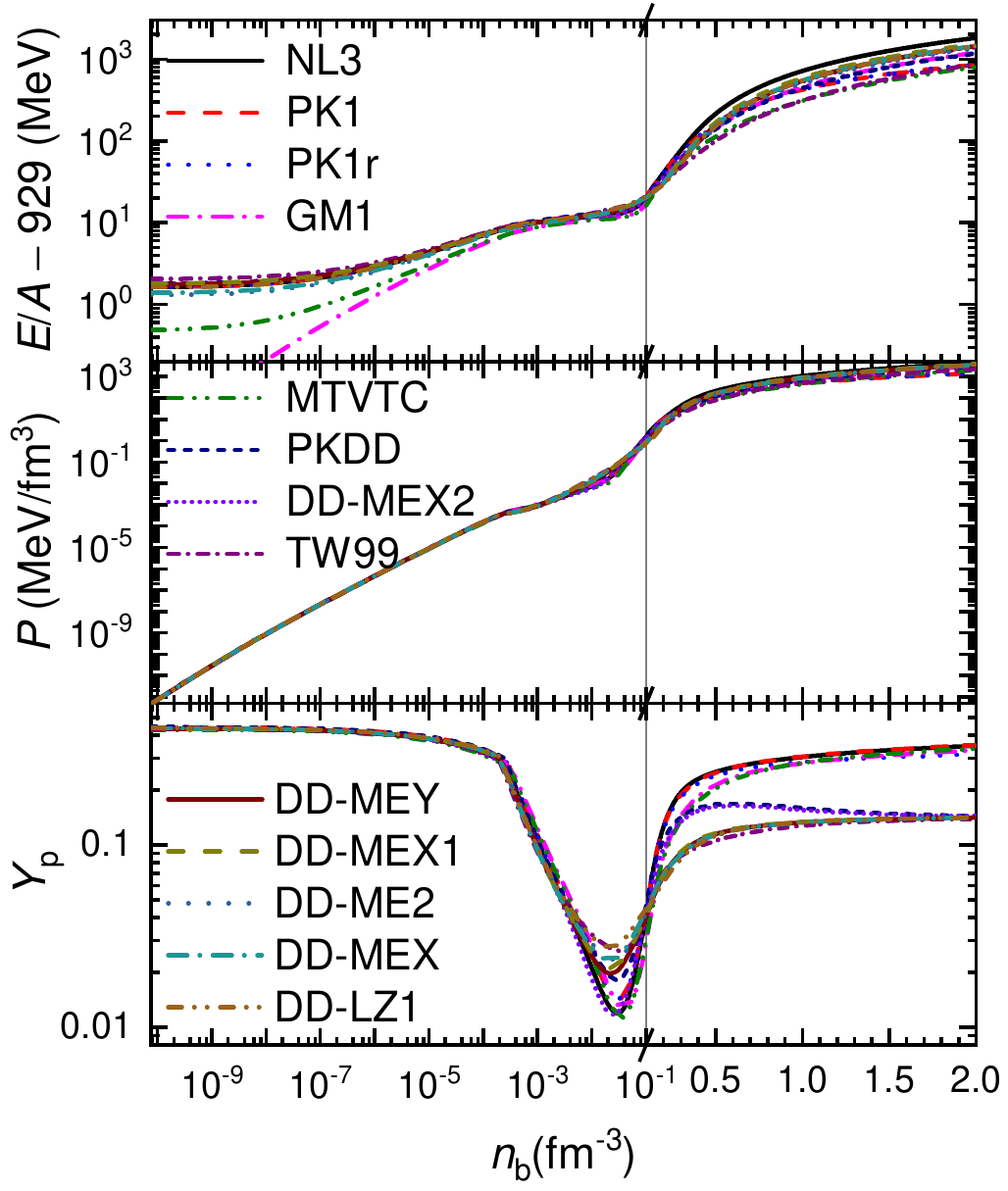}
  \caption{\label{Fig:EOS_beta} Energy per baryon $E/A$, pressure $P$, and proton fraction $Y_p$ of neutron star matter as functions of baryon number density $n_\mathrm{b}$, which are obtained with various RDFs indicated in Table~\ref{table:param}. }
\end{figure}

To fix the EOSs of neutron star matter, based on the results indicated in Sec.~\ref{sec:res_2D_DDLZ1} and \ref{sec:res_2D_All}, we can then fix the proton fractions $Y_p$ at a given baryon number density $n_\text{b}$ by fulfilling the $\beta$ equilibrium condition, i.e.,
\begin{equation}
  \mu_n =\mu_p+\mu_e=\mu_p+\mu_\mu. \label{eq:bstb}
\end{equation}
The obtained proton fractions corresponding to various RDFs are then presented in the bottom panel of Fig.~\ref{Fig:EOS_beta}, while the corresponding energy per baryon and pressure as functions of baryon number density $n_\mathrm{b}$ are also indicated in the top and center panels of Fig.~\ref{Fig:EOS_beta}. At densities $n_\mathrm{b} \lesssim 0.003$ fm${}^{-3}$, the proton fractions predicted by various RDFs generally coincide with each other with slight deviations by up to 0.015. The corresponding binding energy per baryon and pressure of neutron star matter at $n_\mathrm{b} \lesssim 0.003$ fm${}^{-3}$ also generally coincide with each other adopting various RDFs. Note that the variation of the energy per baryon at vanishing densities is generally attributed to the adopted nucleon masses indicated in Table~\ref{table:param}.

As density increases, the proton fractions $Y_p$ decrease while the slopes suddenly change at $Y_p<Y_p^\mathrm{drip}$ ($n_\mathrm{b} \gtrsim 0.0002$ fm${}^{-3}$) with the onset of neutron drip, forming the inner crusts of neutron stars. Then minimum values of $Y_p$ are obtained near $n_\text{b}\approx0.026\ \mathrm{fm}^{-3}$ for neutron star matter. The uncertainty of pressure and energy per baryon due to the variations of RDFs thus increases with density for neutron star matter in the inner crusts. Particularly, at densities within $0.003 \lesssim n_\mathrm{b} \lesssim 0.1$ fm${}^{-3}$, the uncertainty attributed to the variations of RDFs rises and peaks at $n_\mathrm{b} \approx 0.02$ fm${}^{-3}$, which also leads to the variations of the EOSs for neutron star matter.

At larger densities with $n_\text{b} \gtrsim 0.1\ \mathrm{fm}^{-3}$, nuclear pasta becomes unstable and neutron star matter is restored into the uniform phase forming the cores of neutron stars, where muons eventually appear at $n_\text{b}>0.11\ \mathrm{fm}^{-3}$. The uncertainty caused by adopting various RDFs thus grows significantly at larger densities. Note that in this work we have adopted the RDFs that predict the maximum masses of neutron stars surpassing the two-solar-mass limit, while some RDFs predict too large radii and tidal deformations for neutron stars~\cite{Xia2022_CTP74-095303, Huang2024_NPR41-839}. The proton fraction $Y_p$ increases quickly with density, which could eventually trigger the direct Urca processes $n \rightarrow p+e^{-}+\bar \nu_{e}$ and $p+e^{-}\rightarrow n+\nu_{e}$ in neutron stars' cores with $Y_p>0.148$~\cite{Lattimer1991_PRL66-2701}. In particular, the direct Urca processes will take place for neutron stars predicted by NL3, PK1, PK1r, GM1, MTVTC, PKDD, and DD-MEX2 with the slope of symmetry energy $L\gtrsim 80$ MeV, leading to fast cooling in neutron stars. For the RDFs TW99, DD-MEY, DD-MEX1, DD-ME2, DD-MEX, and DD-LZ1, the direct Urca processes never take place in neutron stars' cores.

\begin{figure}
  \includegraphics[width=0.9\linewidth]{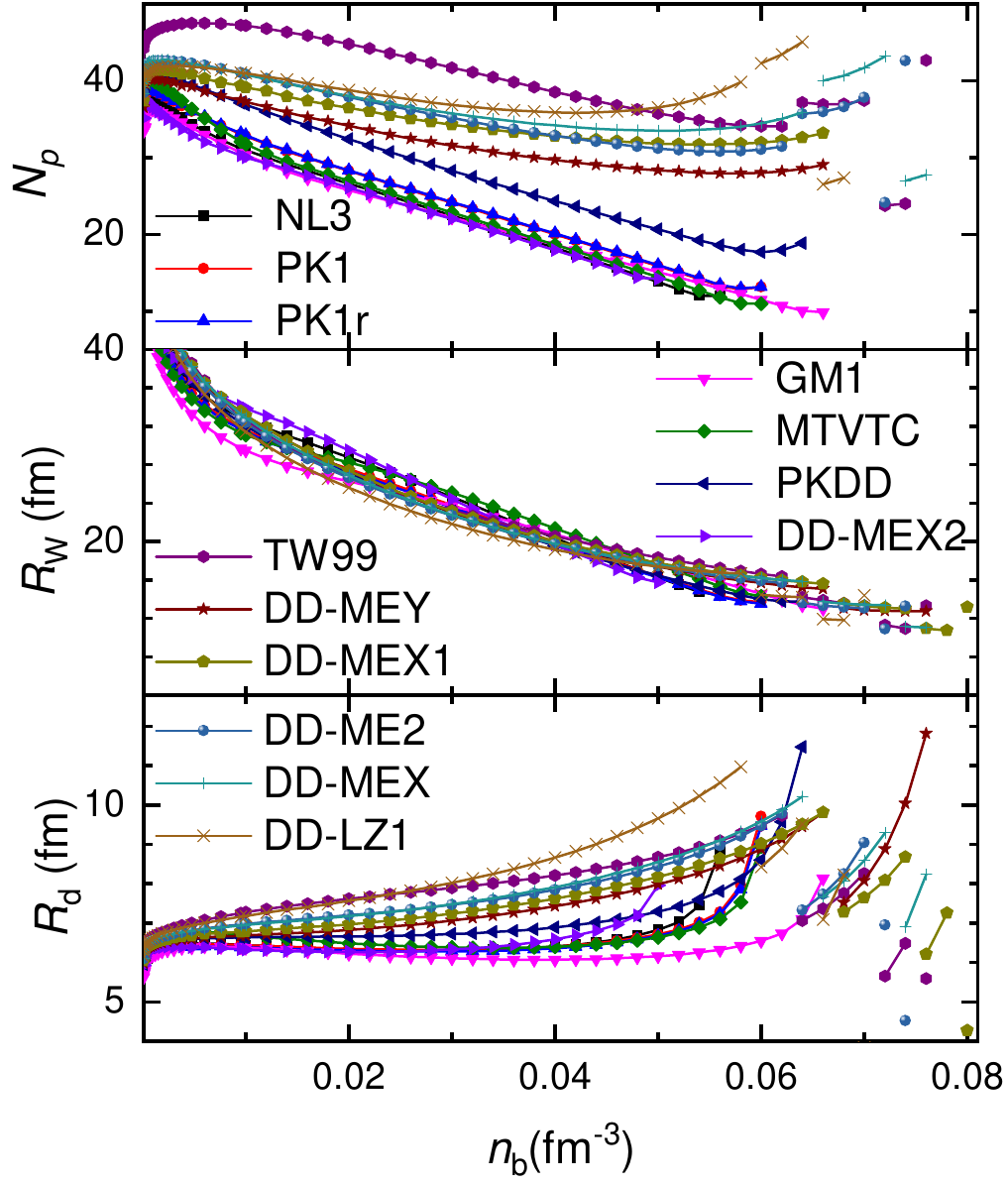}
  \caption{\label{Fig:Mic_beta} Similar to Fig.~\ref{fig:Micro} but for the nonuniform structures in neutron stars' crusts, where the corresponding EOSs are indicated in Fig.~\ref{Fig:EOS_beta}.}
\end{figure}

Beside the EOSs of neutron star matter, the corresponding microscopic structures also play important roles in determining the transport and elastic properties, which affect various phenomenons observed in neutron stars~\cite{Chamel2008_LRR11-10, Caplan2017_RMP89-041002, Zhu2023_PRD107-83023, Sotani2024_Universe10-231}. In correspondence to the EOSs indicated in Fig.~\ref{Fig:EOS_beta}, we then present the microscopic structures of neutron star matter in Fig.~\ref{Fig:Mic_beta}, where the proton number $N_p$, WS cell radius $R_\mathrm{W}$, and droplet size $R_\mathrm{d}$ of neutron star crusts are indicated. Similar to the cases with fixed proton fractions, the droplet, rod, slab, tube, and uniform phases generally appear sequentially as density increases, while the bubble phase does not emerge. In particular, there are only droplet phase if we adopt the RDFs NL3, PK1, PK1r, GM1, MTVTC, PKDD, and DD-MEX2 with the slope of symmetry energy $L\gtrsim 80$ MeV, while the corresponding core-crust transition densities become smaller as well. When we adopt RDFs with smaller $L$, the non-spherical structures such as rod, slab, and tube phases eventually emerge in neutron star crusts. In general, we find that the proton numbers of nuclei, the core-crust transition density, and the onset density of non-spherical nuclei decrease with $L$ for neutron star matter, while the droplet sizes increase. Meanwhile, the WS cell sizes $R_\mathrm{W}$ are roughly insensitive to the adopted RDFs. These findings are generally consistent with previous investigations~\cite{Oyamatsu2007_PRC75-015801, Xu2009_ApJ697-1549, Grill2012_PRC85-055808, Bao2015_PRC91-015807, Shen2020_ApJ891-148, Xia2021_PRC103-055812, Xia2022_CTP74-095303}.

\begin{figure}
\includegraphics[width=\linewidth]{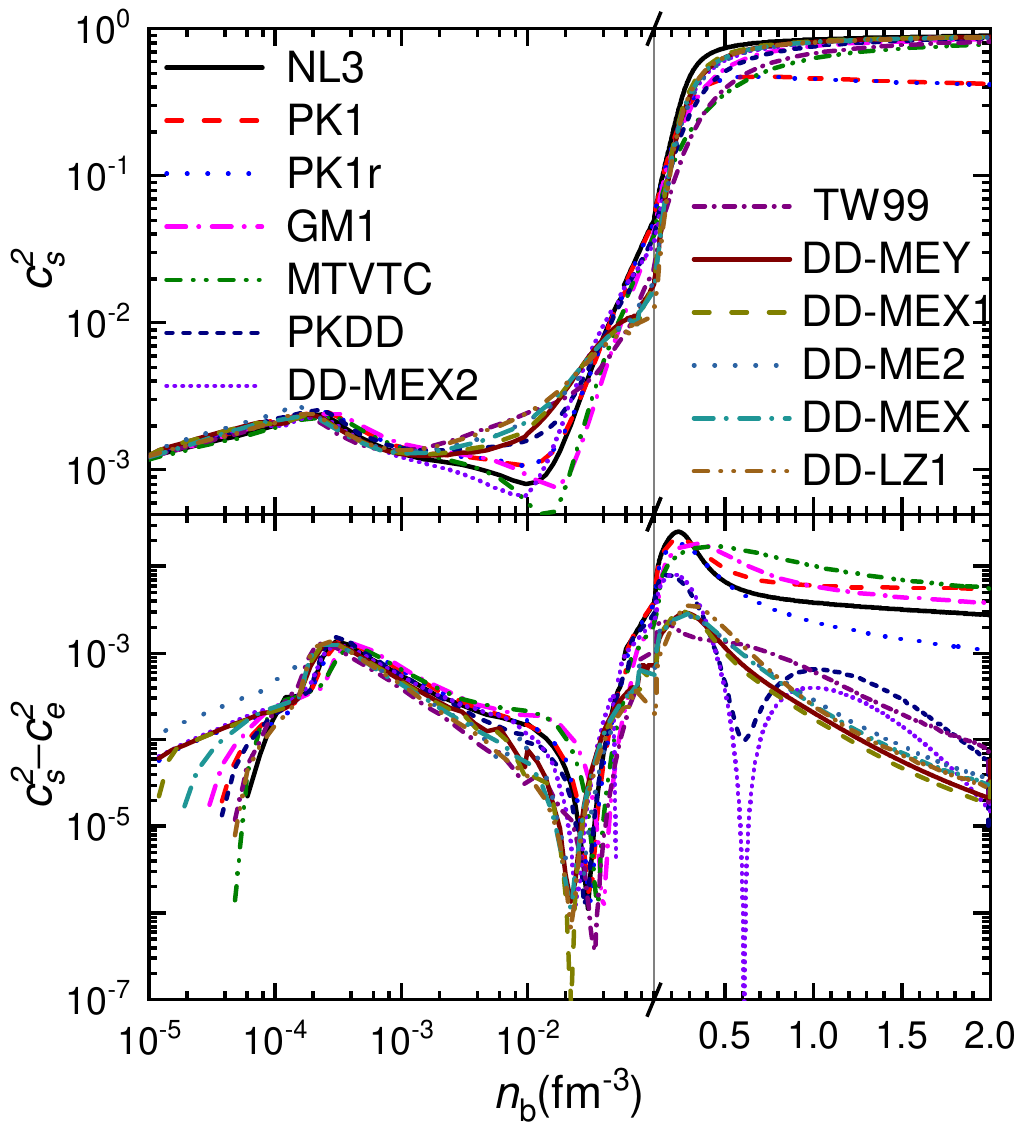}
\caption{\label{Fig:vsound_beta} Adiabatic sound velocity $c^2_s$ and its differences with equilibrium sound velocity $c^2_e$ for neutron star matter as functions of baryon number density $n_\mathrm{b}$, where the corresponding EOSs are indicated in Fig.~\ref{Fig:EOS_beta}.}
\end{figure}

Finally, in Fig.~\ref{Fig:vsound_beta} we present the adiabatic sound velocity $c^2_s$ and its differences with equilibrium sound velocity $c^2_e$ for neutron star matter. The adiabatic and equilibrium sound velocities are fixed by taking derivative of the pressure $P$ with respect to energy density $\varepsilon=E/V$ either holding all the particle fractions $Y_i=n_i/n_\mathrm{b}$ $(i =n, p, e, \mu)$ fixed~\cite{Jaikumar2021_PRD103-123009, Zheng2023_PRD107-103048} or along the $\beta$-stability line indicated in the bottom panel of Fig.~\ref{Fig:EOS_beta}, i.e.,
\begin{equation}
    c^2_e =  \frac{\mbox{d}P}{\mbox{d}\varepsilon}, \ \ \ c^2_s = \left.\frac{\mbox{d}P}{\mbox{d}\varepsilon}\right|_{\{Y_i\}}.    \label{eq:cse}
\end{equation}
The adiabatic index $\Gamma$ is then connected to the adiabatic sound velocity $c^2_s$ via
\begin{equation}
    \Gamma = (\varepsilon/P +1) c^2_s.    \label{eq:Gamma}
\end{equation}
In general, as indicated in Fig.~\ref{Fig:vsound_beta}, the sound velocity is increasing with density, which nevertheless decreases for neutron star matter in the inner crust region above neutron drip densities.

The differences between the two sound velocities $c^2_s-c^2_e$ are nonzero roughly above neutron drip densities, which are caused by the variations of nuclear species/sizes and the emergence of new degrees of freedom, e.g., formation of neutron gas outside of the nuclei above neutron drip densities ($n_\mathrm{b} \approx 0.0002\ \mathrm{fm}^{-3}$),  core-crust transition ($n_\mathrm{b} \approx 0.054$-$0.077\ \mathrm{fm}^{-3}$), and onset of muons ($n_\mathrm{b} \approx 0.105$-$0.128\ \mathrm{fm}^{-3}$). As illustrated in previous investigations~\cite{Jaikumar2021_PRD103-123009, Zheng2023_PRD107-103048, Tran2023_PRC108-015803, Sun2025_PRD111-103019}, nonzero differences in the sound velocities $c^2_s-c^2_e$ will lead to $g$-mode oscillations of neutron stars, which provides opportunities for constraining the EOSs of neutron star matter using various gravitational wave detectors. In particular, the obtained values for $c^2_s-c^2_e$ are nonzero in neutron star crusts, leading to nonzero crust $g$ mode oscillations that was neglected in earlier studies~\cite{Sun2025_PRD111-103019}. The peak value for the differences $c^2_s-c^2_e$ grows with the slope of symmetry energy $L$, causing the frequencies of the core $g$ modes increase. For neutron stars with $M\gtrsim 1$ $M_{\odot}$, the $g$ mode frequencies increase linearly with $L$, offering the possible measurement of $L$ based on the future observations of $g$-mode oscillations~\cite{Sun2025_PRD111-103019}.

\begin{table}
    \caption{\label{tab:number2}Data point numbers and range of $n_\text{b}$ for each EOS data table (attached files ``{RDF\_beta.dat}") of neutron star matter predicted by various RDFs. }
    \begin{tabular}{c|ccc}
    \hline\hline
     Parameter & Range & Step & Total \\
     \hline
     $\text{log}_{10}(n_\mathrm{b}/\mathrm{fm^{-3}})$& $-10.12$ to $-2.02$& 0.1 &  82 \\
      $n_\mathrm{b}/\mathrm{fm^{-3}}$ & 0.01 to 2  & 0.002 &  996 \\
     \hline
    \end{tabular}
\end{table}

Similar as cold dense stellar matter, we then list the obtained results for neutron star matter presented in Figs.~\ref{Fig:EOS_beta}-\ref{Fig:vsound_beta} as data tables illustrated in Table~\ref{tab:number2}.

\section{\label{sec:con}Conclusion}

In this work, including the corresponding microscopic structures, we present a set of EOS tables for dense stellar matter with proton fractions $Y_p =0$-$0.65$ and baryon number densities $n_\text{b}=10^{-8}$-$2 \ \mathrm{fm}^{-3}$ adopting 13 different relativistic density functionals, i.e., NL3~\cite{Lalazissis1997_PRC55-540}, PK1~\cite{Long2004_PRC69-034319},  PK1r~\cite{Long2004_PRC69-034319}, GM1~\cite{Glendenning1991_PRL67-2414}, MTVTC~\cite{Maruyama2005_PRC72-015802}, DD-LZ1~\cite{Wei2020_CPC44-074107}, PKDD~\cite{Long2004_PRC69-034319}, DD-ME2~\cite{Lalazissis2005_PRC71-024312}, TW99~\cite{Typel1999_NPA656-331}, DD-MEX~\cite{Taninah2020_PLB800-135065}, DD-MEX1, DD-MEX2, and DD-MEY~\cite{Taninah2023_PRC107-041301}. The EOSs of dense stellar matter inside neutron stars with baryon number densities $n_\text{b}=7.6\times 10^{-11}$-$2 \ \mathrm{fm}^{-3}$ are obtained as well fulfilling the $\beta$-stability condition. The nonuniform structures (droplet, rod, slab, tube, and bubble) of dense stellar matter are fixed by minimizing the energy per baryon at a given average baryon number density $n_\mathrm{b}$ and proton fraction $Y_p$ in the framework of Thomas-Fermi approximation, where the spherical and cylindrical approximations for the Wigner-Seitz cells are employed. The charge screening effects including contributions from electrons and muons are accounted for with the density profiles altered by the Coulomb potential.

In general, the dense stellar matter exhibits droplet phase at $n_\mathrm{b}\lesssim 0.015\ \mathrm{fm}^{-3}$, while more exotic structures such as rods, slabs, tubes, and bubbles appear sequentially as density increases. At $n_\mathrm{b}\gtrsim0.1 \ \mathrm{fm}^{-3}$, the dense stellar matter become uniform and muons eventually appear. The critical proton fractions $Y_p^\mathrm{drip}$ ($\approx 0.26$-0.31) for neutron drip are obtained, which are roughly insensitive to the adopted RDFs. It is found that the transition densities between different phases vary with $Y_p$ and the adopted RDFs. The proton numbers $N_p$ and droplet sizes $R_\mathrm{d}$ for nuclei in droplet phase generally increase with density $n_\mathrm{b}$ and reach their maximum values at $Y_p\approx Y_p^\mathrm{drip}$, while the WS cell size $R_\mathrm{W}$ is decreasing with $Y_p$ and $n_\mathrm{b}$. Comparing with different nonuniform structures, the droplet sizes $R_\mathrm{d}$ decrease in the order of droplets/bubbles, rods/tubes, and slabs. For the EOSs of dense stellar matter, the pressure $P$ and energy per baryon $E/A$ are increasing with density $n_\mathrm{b}$, while the variation of $E/A$  with respect to $Y_p$ is nonmonotonic. In particular, a minimum $E/A$ at fixed $n_\mathrm{b}$ is observed, which fulfills the $\beta$-equilibrium condition and is typically found in neutron stars. At densities $n_\text{b}\lesssim 0.2 \ \mathrm{fm}^{-3}$, the pressure $P$ increases with $Y_p$ while at larger densities the variation becomes less distinctive.

For dense stellar matter at small densities ($n_\text{b}\lesssim 10^{-5} \ \mathrm{fm}^{-3}$) or large proton fractions ($n_\text{b}\lesssim0.1 \ \mathrm{fm}^{-3}$ and $Y_p\gtrsim Y_p^\mathrm{drip}$), the EOSs and microscopic structures are generally insensitive to the adopted RDFs. The variations due to the uncertainties in RDFs emerge at $10^{-5} \lesssim n_\text{b}\lesssim0.1 \ \mathrm{fm}^{-3}$ and $Y_p\lesssim Y_p^\mathrm{drip}$ with the onset of neutron drip. In particular, the obtained values of $N_p$ and $R_\mathrm{d}$ approximately decrease with $L$ at $Y_p\lesssim Y_p^\mathrm{drip}$, while the values of $R_\mathrm{W}$ also decreases slightly. If RDFs with smaller $L$ are adopted, more nuclear pasta structures emerge and the corresponding density range becomes larger at $Y_p\lesssim Y_p^\mathrm{drip}$. At larger densities with $n_\text{b}\gtrsim 0.1 \ \mathrm{fm}^{-3}$, the dense stellar matter becomes uniform and the uncertainties in the EOSs grow significantly.

For dense stellar matter inside neutron stars, the EOSs and microscopic structures are generally insensitive to the adopted RDFs at  $n_\text{b}\lesssim 10^{-4} \ \mathrm{fm}^{-3}$. Above the neutron drip density ($n_\mathrm{b} \approx 0.0002\ \mathrm{fm}^{-3}$), similar to the cases with fixed proton fractions at $Y_p\lesssim Y_p^\mathrm{drip}$, the uncertainties in the EOSs and microscopic structures grow. The non-spherical structures such as rod, slab, and tube phases emerge only when we adopt RDFs with small $L\lesssim 60$ MeV. Meanwhile, the proton numbers of nuclei, the core-crust transition density, and the onset density of non-spherical nuclei decrease with $L$, while the droplet sizes increase. Nevertheless, the WS cell sizes $R_\mathrm{W}$ are roughly insensitive to the adopted RDFs. The differences between the adiabatic and equilibrium sound velocities $c^2_s-c^2_e$ for dense stellar matter in the inner crusts of neutron stars are sizable, which are caused by the variations of nuclear species/sizes and emergence of neutron gas outside of nuclei above neutron drip densities. This leads to nonzero crust $g$ mode oscillations that was neglected in earlier studies~\cite{Sun2025_PRD111-103019}. At larger densities with $n_\text{b} \gtrsim 0.1\ \mathrm{fm}^{-3}$, neutron star matter becomes uniform and muons eventually appear at $n_\text{b}>0.11\ \mathrm{fm}^{-3}$, where the uncertainties caused by adopting various RDFs grow significantly. For example, the peak value for the differences $c^2_s-c^2_e$ grows with the slope of symmetry energy $L$, so that the frequencies of the core $g$ modes in neutron stars increase, offering the possible measurement of $L$ based on the future observations of $g$-mode oscillations~\cite{Sun2025_PRD111-103019}.

\section*{ACKNOWLEDGMENTS}
This work was supported by the National Natural Science Foundation of China (Grant No. 12275234) and the National SKA Program of China (Grant No. 2020SKA0120300).

\newpage
%

\end{document}